\documentclass[12,preprint,aps,prd,nofootinbib,superscriptaddress, tightenlines]{revtex4}

\usepackage[OT1]{fontenc}
\usepackage[utf8]{inputenc}
\usepackage[dvipsnames]{xcolor}
\usepackage{graphics}
\usepackage{lscape}
\usepackage{rotfloat}
\usepackage{rotating}
\usepackage{amsmath}
\usepackage{amssymb}
\usepackage{graphicx}                                                                        
\usepackage{flexisym}
\usepackage{breqn}
\usepackage{mathtools}
\usepackage{slashed}
\usepackage{hyperref}
\usepackage{enumitem}

\newcommand{\bdm}{\begin{dmath}}
\newcommand{\edm}{\end{dmath}}
\newcommand{\bdms}{\begin{dmath*}}
\newcommand{\edms}{\end{dmath*}}
\newcommand{\bdg}{\begin{dgroup*}}
\newcommand{\edg}{\end{dgroup*}}  
\newcommand{\be}{\begin{equation}}
\newcommand{\ee}{\end{equation}}
\newcommand{\bea}{\begin{eqnarray}} 
\newcommand{\eea}{\end{eqnarray}}
\def\openone{\leavevmode\hbox{\small1\kern-3.3pt\normalsize1}}

\newcommand{\MSbar}{\overline{\textrm{MS}}} 

\definecolor{mygreen}{RGB}{0,153,0}

\definecolor{myorange}{RGB}{255,130,0}

\begin{document}
%%%%%%%%%%%%%%%%%%%%%%%%%%%%%%%%%%%%%%%%%%%%%%%%%%%%%%%%%%%%%%%%%%%%%%%%%%%%%
\title{One-loop renormalization of staple-shaped operators in continuum and lattice regularizations}
\author{Martha Constantinou}
\email{marthac@temple.edu}
\affiliation{$\it{Department \ of \ Physics, \ Temple \ University, \ Philadelphia, \ Pennsylvania \ 19122 - 1801, \ USA}$}
\author{Haralambos Panagopoulos} 
\email{haris@ucy.ac.cy}
\affiliation{$\it{Department \ of \ Physics, \ University \ of \ Cyprus, \ POB \ 20537, \ 1678, \ Nicosia, \ Cyprus \qquad}$ \vspace{0.5cm}}
\author{Gregoris Spanoudes \vspace{0.5cm}} 
\email{spanoudes.gregoris@ucy.ac.cy}
\affiliation{$\it{Department \ of \ Physics, \ University \ of \ Cyprus, \ POB \ 20537, \ 1678, \ Nicosia, \ Cyprus \qquad}$ \vspace{0.5cm}}
%----------------------------------------------------------------------------
\begin{abstract}
\vspace{0.5cm} In this paper we present one-loop results for the renormalization of nonlocal quark bilinear operators, containing a staple-shaped Wilson line, in both continuum  and lattice regularizations. The continuum calculations were performed in dimensional regularization, and the lattice calculations for the Wilson/clover fermion action and for a variety of Symanzik-improved gauge actions.  We extract the strength of the one-loop linear and logarithmic divergences (including cusp divergences), which appear in such nonlocal operators; we identify the mixing pairs which occur among some of these operators on the lattice, and we calculate the corresponding mixing coefficients. We also provide the appropriate RI$'$-like scheme, which disentangles this mixing nonperturbatively from lattice simulation data, as well as the one-loop expressions of the conversion factors, which turn the lattice data to the $\MSbar$ scheme. Our results can be immediately used for improving recent nonperturbative investigations of transverse momentum-dependent distribution functions on the lattice. 

Finally, extending our perturbative study to general Wilson-line lattice operators with $n$ cusps, we present results for their renormalization factors, including identification of mixing and determination of the corresponding mixing coefficients, based on our results for the staple operators.
\end{abstract}
%----------------------------------------------------------------------------
\hspace{-5mm}\begin{minipage}{\textwidth}
\maketitle
\end{minipage}
\renewcommand{\thefootnote}{\alph{footnote}}
\footnotetext{Electronic addresses: ${}^{a}$ marthac@temple.edu, \ ${}^{b}$ haris@ucy.ac.cy, \ ${}^{c}$ spanoudes.gregoris@ucy.ac.cy}
\renewcommand{\thefootnote}{\arabic{footnote}}
%----------------------------------------------------------------------------
\section{Introduction}
\label{sec:intro}

One of the main research directions of nuclear and particle physics is the study of the rich internal structure of hadrons, which are the building blocks of the visible Universe. Quantum chromodynamics (QCD) is the theory governing the strong interactions, which are responsible for binding partons (quark and gluons) together into hadrons. Despite the various theoretical models that have been developed for the investigation of hadron structure (e.g., diquark spectator and chiral quark models), an \textit{ab initio} calculation is desirable to capture the full QCD dynamics. Due to the complexity of the QCD Lagrangian, an analytic solution is not possible, and numerical simulations (lattice QCD) may be used as a \textit{first principle} formulation to study the properties of fundamental particles.

\bigskip

Distribution functions consist of a set of nonperturbative quantities that describe hadron structure and have the advantage of being process-independent and accessible both experimentally and theoretically. They are expressed in terms of variables defined in the longitudinal and transverse directions with respect to the hadron momentum. Based on this, the distribution functions may be classified into parton distribution functions (PDFs), generalized parton distributions (GPDs) and transverse-momentum dependent parton distribution functions (TMDs). Important information is still missing for all three types of distributions: The most well-studied are PDFs, which are single-variable functions, while the TMDs are only very limitedly studied due to the difficulty in extracting them experimentally and theoretically. However, TMDs are crucial for the complete understanding of hadron structure as they complement, together with GPDs, the three-dimensional picture of a hadron. 

\bigskip 

Due to their light cone nature, distribution functions cannot be computed directly on a Euclidean lattice and typically are parametrized in terms of local operators that give their moments. The distribution functions can thus be recovered from an operator product expansion (OPE), which is, however, a very difficult task: Signal-to-noise ratio decreases with the addition of covariant derivatives in the operators, and an unavoidable power-law mixing under renormalization appears for higher moments. Nevertheless, information on distribution functions (mainly PDFs and, to a lesser extent GPDs) from lattice QCD was obtained from their first moments, via calculations of matrix elements of local operators. These moments are directly related to measurable quantities, for example, the axial charge and quark momentum fraction. 

\bigskip

Novel approaches for an \textit{ab initio} evaluation of distribution functions on the lattice have been employed in recent years. In these approaches, nonlocal operators, including a Wilson line, are involved. While local operators have been used extensively in perturbative and nonperturbative calculations, nonlocal operators were limitedly studied. In particular, calculations using nonlocal operators with Wilson lines in a variety of shapes appear in the literature within continuum perturbation theory. Starting from the seminal work of Mandelstam \cite{Mandelstam:1968hz}, Polyakov \cite{Polyakov:1979gp}, Makeenko-Migdal \cite{Makeenko:1979pb}, there have been investigations of the renormalization of Wilson loops for both smooth \cite{Dotsenko:1979wb} and nonsmooth \cite{Brandt:1981kf} contours. Due to the presence of the Wilson line, power-law divergences arise for cutoff regularized theories, such as lattice QCD. It has been proven that in the case of dimensional regularization and in the absence of cusps and self-intersections, all divergences in Wilson loops can be reabsorbed into a renormalization of the coupling constant~\cite{Dotsenko:1979wb}. Wilson-line operators have been studied with a number of approaches, including an auxiliary-field formulation~\cite{Craigie:1980qs,Dorn:1986dt}, and the Mandelstam formulation \cite{Stefanis:1983ke}. Particular studies of Wilson-line operators with cusps in one and two loops, can be found in Refs. \cite{Brandt:1981kf, Knauss:1984rx, Korchemsky:1987wg}. There is also related work, in the context of the heavy quark effective theory (HQET)\footnote{The interrelation between Wilson-line operators and HQET currents is demonstrated in Ref. \cite{Korchemsky:1991zp}.} \cite{Shifman:1986sm, Politzer:1988wp, Falk:1990yz, Ji:1991az}, including investigations in three loops \cite{Chetyrkin:2003vi}. 

\bigskip

Computations of matrix elements using nonlocal operators with a straight Wilson line have been revived in lattice QCD and phenomenology mainly due to their connection to PDFs via the quasi-PDFs approach proposed by X. Ji~\cite{Ji:2013dva}\footnote{The same operators are used in an alternative approach (pseudo-PDFs) to extract light cone PDFs ~\cite{Radyushkin:2016hsy,Radyushkin:2017ffo,Radyushkin:2017cyf}. Earlier ideas for accessing $x$-dependent PDFs are summarized in Ref.~\cite{Cichy:2018mum}.}. Several aspects of the properties of nonlocal Wilson-line operators have been addressed, including the feasibility of a calculation from lattice QCD~\cite{Lin:2014zya,Alexandrou:2015rja,Chen:2016utp,Alexandrou:2016jqi}, their renormalizability~\cite{Ji:2015jwa,Ishikawa:2017faj,Ji:2017oey,Wang:2017eel,Xiong:2017jtn,Zhang:2018diq,Li:2018tpe} and appropriate renormalization prescriptions~\cite{Constantinou:2017sej,Alexandrou:2017huk,Spanoudes:2018zya}. The renormalization has proven to be a challenging and delicate process in which a number of new features emerge, as compared to the case of local operators: There appears an additional power-law divergence, and the matrix elements are nonlocal and contain an imaginary part.

\bigskip

While information on physical quantities is obtained from hadron matrix elements, calculated nonperturbatively in numerical simulations of lattice QCD, perturbation theory has played a crucial role in the development of a complete renormalization prescription based on Ref.~\cite{Constantinou:2017sej}. In the latter work the renormalization was addressed in lattice perturbation theory and a finite mixing was identified among nonlocal operators of twist-2 and twist-3. The complete mixing pattern discussed in Refs.~\cite{Constantinou:2017sej} led to the proposal of a nonperturbative RI-type scheme~\cite{GHP,Alexandrou:2017huk}, also employed in Ref.~\cite{Chen:2017mzz}. This development of the renormalization of nonlocal operators has been a crucial aspect in state-of-the-art numerical simulations, e.g. the work of Refs.~\cite{Alexandrou:2018pbm,Alexandrou:2018eet} (for a recent overview on lattice QCD calculations see Ref.~\cite{Cichy:2018mum} and references therein).

\bigskip

In this work we generalize the calculation of Ref.~\cite{Constantinou:2017sej} to include nonlocal operators with a staple-shaped Wilson line. We compute their Green's functions to one-loop level in perturbation theory using dimensional (DR) and lattice (LR) regularizations. The functional form of the Green's functions reveals the renormalization pattern and mixing among operators of different Dirac structure, in each regularization. We find that these operators renormalize multiplicatively in DR, but have finite mixing in LR. Results for both regularizations have been combined to extract the renormalization functions in the lattice $\MSbar$ scheme. In addition, the results in DR have been used to obtain the conversion factor between RI-type and $\MSbar$ schemes. We also present an extension to operators containing a Wilson line of arbitrary shape on the lattice, with $n$ cusps. Preliminary results of the current work have been presented in Ref. \cite{Spanoudes:2018jtc}.

\bigskip

\begin{figure}[!ht]
\includegraphics[scale=0.9]{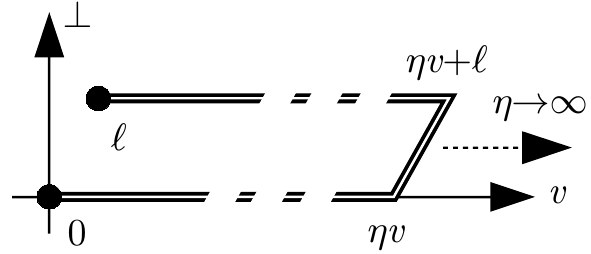}
\caption{Staple-shaped gauge links as used in analyses of SIDIS and Drell-Yan processes. For notation, see Ref.~\cite{Engelhardt:2015xja}.}
\label{fig:staple}
\end{figure}

Staple-shaped nonlocal operators (see Fig.~\ref{fig:staple}) are crucial in studies of TMDs, which encode important details on the internal structure of hadrons. In particular, they give access to the intrinsic motion of partons with respect to the transverse momentum, through the formalism of QCD factorization, that can be used to link experimental data to the three-dimensional partonic structure of hadrons. An operator with a staple of infinite length, $\eta{\to}\infty$, (see Fig.~\ref{fig:staple}) enters the analysis of semi-inclusive deep inelastic scattering (SIDIS) processes\footnote{Staple-shaped operators appear also in Drell-Yan process, with the staple oriented in the opposite direction compared to SIDIS.} in a kinematical region where the photon virtuality is large and the measured transverse momentum of the produced hadron is of the order of $\Lambda_{\rm QCD}$~\cite{Bomhof:2006dp}. 

\bigskip

To date, only limited studies of TMDs exist in lattice QCD (see, e.g., Refs.~\cite{Musch:2010ka,Musch:2011er,Engelhardt:2015xja,Yoon:2017qzo} and references therein), such as the generalized Sivers and Boer-Mulders transverse momentum shifts for the SIDIS and Drell-Yan cases. These studies include staple links of finite length that is restricted by the spatial extent of the lattice volume. To recover the desired infinite length one checks for convergence as the length increases, and an extrapolation to $\eta{\to}\infty$ is applied. More recently, the connection between nonlocal operators with staple-shaped Wilson line and orbital angular momentum~\cite{Engelhardt:2017miy,Raja:2017xlo} has been discussed. This relies on a comparison between straight and staple-shaped Wilson lines, with the staple-shaped path yielding the Jaffe-Manohar~\cite{Hatta:2011ku,Burkardt:2012sd} definition of quark orbital angular momentum, and the straight path yielding Ji's definition~\cite{Burkardt:2012sd,Ji:2012sj,Rajan:2016tlg}. The difference between these two can be understood as the torque experienced by the struck quark as a result of final state interactions~\cite{Burkardt:2012sd,Ji:2012sj}.

\bigskip

An important aspect of calculations in lattice QCD is the renormalization that needs to be applied on the operators under study (unless conserved currents are used). As is known from older studies~\cite{Dotsenko:1979wb,Arefeva:1980zd,Craigie:1980qs,Dorn:1986dt,Stefanis:1983ke}, the renormalization of Wilson-line operators in continuum theory (except DR) includes a divergent term $e^{-\delta m L}$, where $\delta m$ is a dimensionful quantity whose magnitude diverges linearly with the regulator, and $L$ is the total length of the contour. For staple-shaped operators, $L = (2 |y| + |z|)$, where $y \lessgtr 0$ and $z \lessgtr 0$ define the extension of the staple in the $y{-}z$ plane, chosen to be spatial. The existing lattice calculations of staple-shaped operators assume that the lattice operators have the same renormalization properties as the continuum operators, in particular that there is no mixing present. This allows one to focus on ratios between such operators~\cite{Musch:2010ka,Musch:2011er,Engelhardt:2015xja,Yoon:2017qzo} in order to cancel multiplicative renormalization, which is currently unknown\footnote{The question of whether nonlocal operators with staple-shaped Wilson lines renormalize multiplicatively was raised in Ref.~\cite{Yoon:2017qzo} after our work on straight Wilson-line operators~\cite{Constantinou:2017sej}.}. However, as we show in this paper, this is not the case for operators where finite mixing is present and must be taken into account.

\bigskip

One of the main goals of this work is to provide important information that may impact nonperturbative studies of TMDs and potentially lead to the development of a nonperturbative renormalization prescription similar to the case of quasi-PDFs discussed above. The paper is organized in five sections including the following: In Sec. \ref{CalculationSetup} we provide the set of operators under study, the lattice formulation, the renormalization prescription for nonlocal operators that mix under renormalization and the basics of the conversion to the $\MSbar$ scheme. Section \ref{Calculationprocedure} presents our main results in dimensional and lattice regularization. This includes both the renormalization functions and conversion factors between the RI$'$ and $\MSbar$ schemes. An extension of the work to include general nonlocal Wilson-line operators with $n$ cusps is presented in Sec. \ref{sec.IV}, while in Sec. \ref{sec.V} we give a summary and future plans. For completeness we include two appendices where we give the expressions for the Green's functions in dimensional regularization (Appendix \ref{ap.A}) as well as the expressions related to the renormalization of the fermion fields (Appendix \ref{ap.B}).

\section{Calculation Setup}
\label{CalculationSetup}
In this section we briefly introduce the setup of our calculation, along with the notation used in the paper. We give the definitions of the operators and the lattice actions; we also provide the renormalization prescriptions that we use in the presence of operator mixing.
\subsection{Operator setup}
\label{operatorsetup}
\begin{figure}[!htbp]
\centering
\includegraphics[scale=0.8]{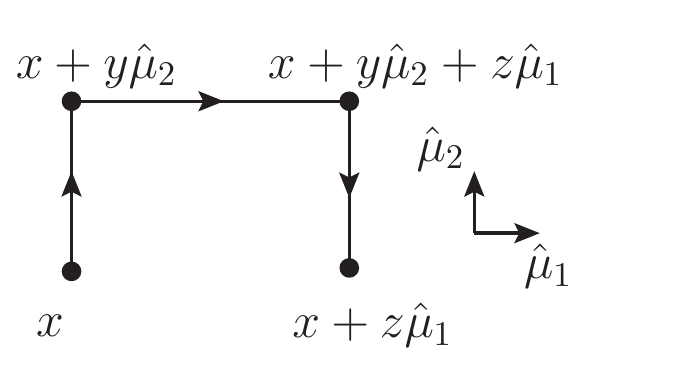}
\caption{Staple-shaped Wilson line $W(x,x+y \hat{\mu}_2,x+y \hat{\mu}_2+z \hat{\mu}_1,x+z \hat{\mu}_1)$.}
\label{fig:Staple2}
\end{figure}

The staple-shaped Wilson-line operators have the following form: 
\begin{gather}
\mathcal{O}_\Gamma \equiv \bar\psi(x) \ \Gamma \ W(x,x+y \hat{\mu}_2,x+y \hat{\mu}_2+z \hat{\mu}_1,x+z \hat{\mu}_1) \ \psi(x + z {\hat{\mu}}_1),  
\end{gather}
where $W$ denotes a staple with side lengths $|z|$ and $|y|$, which lies in the plane specified by the directions $\hat{\mu}_1$ and $\hat{\mu}_2$ (see Fig. \ref{fig:Staple2}); it is defined by
\begin{align}
W(x, x + &y \hat{\mu}_2, x + y \hat{\mu}_2 + z \hat{\mu}_1, x + z \hat{\mu}_1) = \nonumber \\
&\mathcal{P} \Big\{\left(e^{i g \int_0^y d \zeta A_{\mu_2} (x + \zeta {\hat{\mu}}_2)}\right) \cdot \left(e^{i g \int_0^z d \zeta A_{\mu_1} (x + y \hat{\mu}_2 + \zeta \hat{\mu}_1)}\right) \cdot \left( e^{i g \int_0^y d \zeta A_{\mu_2} (x + z {\hat{\mu}}_1 + \zeta {\hat{\mu}}_2)} \right)^\dagger \Big\}. 
\label{stapleW}
\end{align} 
The symbol $\Gamma$ can be one of the following Dirac matrices: $\openone$, $\gamma_5$, $\gamma_\mu$, $\gamma_5 \gamma_\mu$, $\sigma_{\mu \nu}$ (where $\mu, \nu = 1, 2, 3, 4$ and $\sigma_{\mu \nu} = [\gamma_\mu, \gamma_\nu]/2$). For convenience, we adopt the following notation for each Dirac matrix: $S \equiv \openone$, $P \equiv \gamma_5$, $V_\mu \equiv \gamma_\mu$, $A_\mu \equiv \gamma_5 \gamma_\mu$, $T_{\mu \nu} \equiv \sigma_{\mu \nu}$ and the standard nomenclature for the corresponding operators: $\mathcal{O}_S:$ scalar, $\mathcal{O}_P:$ pseudoscalar, $\mathcal{O}_{V_\mu}:$ vector, $\mathcal{O}_{A_\mu}:$ axial-vector and $\mathcal{O}_{T_{\mu \nu}}:$ tensor. Of particular interest is the study of vector, axial-vector and tensor operators, which correspond to the three types of TMDs: unpolarized, helicity and transversity, respectively.

\bigskip

The fermion and antifermion fields appearing in $\mathcal{O}_\Gamma$ can have different flavor indices. Operators with different flavor content cannot mix among themselves; further, for mass-independent renormalization schemes, flavor-nonsinglet operators which differ only in their flavor content will have the same renormalization factors and mixing coefficients. Results for the flavor-singlet case will be identical to those for the flavor-nonsinglet case at one loop, but they will differ beyond one loop and nonperturbatively; however, the setup described below [Eqs. (\ref{OGamma1} -- \ref{LambdaGamma3})] will be identical in both cases.

\subsection{Lattice actions}

In our lattice calculation we make use of the Wilson/clover fermion action \cite{Sheikholeslami:1985ij}. In standard notation it reads

\begin{align}
S_{\rm F} = &- \frac{a^3}{2} \sum_{x, f, \mu} \, \overline{\psi}_f (x) \Big[(r - \gamma_\mu ) U_\mu (x) \ \psi_f (x + a \hat{\mu}) + (r + \gamma_\mu ) U_\mu^\dagger (x - a \hat{\mu}) \ \psi_f (x - a \hat{\mu}) \Big] \nonumber \\
&+ a^3 \sum_{x,f} \, \overline{\psi}_f (x) \ (4r + a m_o^f) \ \psi_f (x) \nonumber \\
&- \frac{a^3}{32} \sum_{x, f, \mu, \nu} c_{SW} \ \overline{\psi}_f (x) \ \sigma_{\mu \nu} \ \Big[ Q_{\mu \nu} (x) - Q_{\nu \mu} (x) \Big] \ \psi_f (x)\,,
\label{WFaction1}
\end{align}
where $a$ is the lattice spacing and
\begin{eqnarray}
Q_{\mu \nu} &=& U_\mu (x) \ U_\nu (x + a \hat{\mu}) \ U_\mu^\dagger (x + a \hat{\nu}) \ U_\nu^\dagger (x) \nonumber \\  
&+& U_\nu (x) \ U_\mu^\dagger (x + a \hat{\nu} - a \hat{\mu}) \ U_\nu^\dagger (x - a \hat{\mu}) \ U_\mu (x - a \hat{\mu}) \nonumber \\ 
&+& U_\mu^\dagger (x - a \hat{\mu}) \ U_\nu^\dagger (x - a \hat{\mu} -a \hat{\nu}) \ U_\mu (x - a \hat{\mu} - a \hat{\nu}) \ U_\nu (x - a \hat{\nu}) \nonumber \\ 
&+& U_\nu^\dagger (x - a \hat{\nu}) \ U_\mu (x - a \hat{\nu}) \ U_\nu (x + a \hat{\mu} - a \hat{\nu}) \ U_\mu^\dagger (x).
\end{eqnarray}
Following common practice, we henceforth set the Wilson parameter r equal to 1. The clover coefficient $c_{SW}$ will be treated as a free parameter, for wider applicability of the results. The mass term ($\sim m_0^f$) will be irrelevant in our one-loop calculations, since we will apply mass-independent renormalization schemes. The above formulation, and thus our results, are also applicable to the twisted mass fermions \cite{Shindler:2007vp} in the massless case. One should, however, keep in mind that, in going from the twisted basis to the physical basis, operator identifications are modified (e.g., the scalar density, under ``maximal twist'', turns into a pseudoscalar density, etc.).

\bigskip

\begin{figure}[!htbp]
\centering
\includegraphics[scale=0.8]{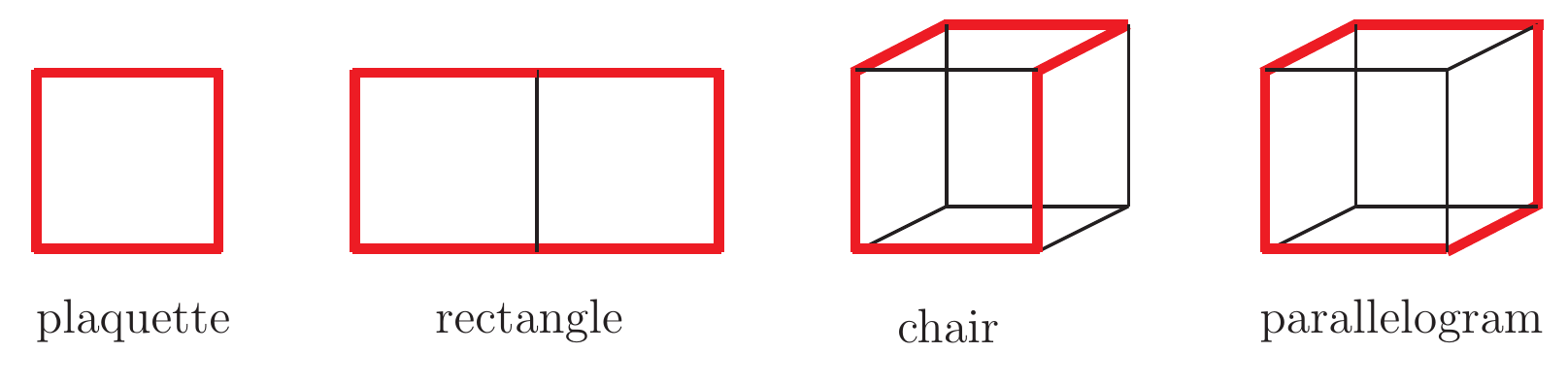}
\caption{The four Wilson loops of the Symanzik improved gauge actions.}
\label{figSym}
\end{figure}

For gluons, we employ a family of Symanzik improved actions \cite{Horsley:2004mx}, of the form,
\bea
\hspace{-1cm}
S_G=\frac{2}{g_0^2} \Bigl[ &c_0& \sum_{\rm plaq.} {\rm Re\,Tr\,}\{1-U_{\rm plaq.}\}
%\nonumber \\ 
\,+\, c_1 \sum_{\rm rect.} {\rm Re \, Tr\,}\{1- U_{\rm rect.}\} 
\nonumber \\ 
+ &c_2& \sum_{\rm chair} {\rm Re\, Tr\,}\{1-U_{\rm chair}\} 
%\nonumber \\ 
\,+\, c_3 \sum_{\rm paral.} {\rm Re \,Tr\,}\{1-U_{\rm paral.}\}\Bigr]\,,
\label{Symanzik}
\eea
where $U_{\rm plaq.}$ is the 4-link Wilson loop and $U_{\rm rect.}$, $U_{\rm chair}$, $U_{\rm paral.}$ are the three possible independent 6-link Wilson loops (see Fig. ~\ref{figSym}). The Symanzik coefficients $c_i$ satisfy the following normalization condition:
\be
c_0 + 8 c_1 + 16 c_2 + 8 c_3 = 1.\,
\label{norm}
\ee

\vspace{-0.5cm} 

\bigskip

For the numerical integration over loop momenta we selected a variety of values for $c_i$; for the sake of compactness, in what follows we will present only results for some of the most frequently used sets of values, corresponding to $c_2 = c_3 = 0$, as shown in Table \ref{tab1}.
\begin{table}[!h]
\centering
\begin{tabular}{|l|l|l|}
\hline
\ \textbf{Gluon action} \ & \ $\boldsymbol{c_0}$ \ & \ $\boldsymbol{c_1}$ \ \\
\hline
\ Wilson \ & \ 1 \ & \ \ 0 \ \\
\ Tree-level Symanzik \ & \ 5/3 \ & \ -1/12 \ \\
\ Iwasaki \ & \  3.648 \ & \ -0.331$\,$ \ \\
\hline
\end{tabular}
\caption{Values of the Symanzik coefficients for selected gluon actions. The coefficients $c_2$ and $c_3$ equal 0 for these actions.}
\label{tab1}
\end{table} 

\subsection{Renormalization prescription}
\label{sec.IIC}
The renormalization of nonlocal operators is a nontrivial process. As shown in our study of straight-line operators in Ref. \cite{Constantinou:2017sej}, a hidden operator mixing is present in chirality-breaking regularizations, such as the Wilson/clover fermions on the lattice. This mixing does not involve any divergent terms; it stems from finite regularization-dependent terms, which are not present in the $\MSbar$ renormalization scheme, as defined in dimensional regularization (DR). Thus, our first goal is to compute perturbatively all renormalization functions and mixing coefficients which arise in going from the lattice regularization (LR) to the $\MSbar$ scheme. Ultimately, a nonperturbative evaluation of all these quantities is desirable; to this end, and given that the very definition of $\MSbar$ is perturbative, we must devise an appropriate, RI$'$-type renormalization prescription which reflects the operator mixing. We will proceed with the definition of the renormalization factors of operators, as mixing matrices, in textbook fashion. We modify our prescription in Ref. \cite{Constantinou:2017sej} to correspond to the resulting operator-mixing pairs of the present calculation, which are different from those found in the straight-line operators. The reason behind this difference is explained in detail in Sec. \ref{sec.IV}. The mixing pairs found from our calculation on the lattice are (see Sec. \ref{sec.Mixing}): $(\mathcal{O}_P, \mathcal{O}_{A_{\mu_2}}), (\mathcal{O}_{V_i}, \mathcal{O}_{T_{i\mu_2}})$, where $i$ can be any of the three orthogonal directions to the $\hat{\mu}_2$ direction. The remaining operators do not show any mixing, and thus their renormalization factors have the typical $1 \times 1$ form [see Eq. \eqref{OGamma3}]. Taking into account all the above, we define the renormalization factors which relate each bare operator $\mathcal{O}_{\Gamma}$ with the corresponding renormalized one via the following equations:
\bea
\begin{pmatrix}
\vspace{0.1cm}\mathcal{O}_P^Y \ \ \\
\mathcal{O}_{A_{\mu_2}}^Y
\end{pmatrix}
\! \! \! &=&
{\begin{pmatrix}
\vspace{0.1cm}Z^{X,Y}_P \quad & Z^{X,Y}_{(P,A_{\mu_2})} \\
Z^{X,Y}_{(A_{\mu_2},P)} & \hspace{-0.45cm}Z^{X,Y}_{A_{\mu_2}}
\end{pmatrix}}^{-1} \
\begin{pmatrix}
\vspace{0.1cm}\mathcal{O}_P \ \ \\
\mathcal{O}_{A_{\mu_2}}
\end{pmatrix},
\label{OGamma1} \\
\begin{pmatrix}
\vspace{0.1cm}\mathcal{O}_{V_i}^Y \ \ \\
\mathcal{O}_{T_{i \mu_2}}^Y
\end{pmatrix}
\! \! \! &=& 
{\begin{pmatrix}
\vspace{0.1cm}Z^{X,Y}_{V_i} \quad & Z^{X,Y}_{(V_i,T_{i \mu_2})} \\
Z^{X,Y}_{(T_{i \mu_2},V_i)} & \hspace{-0.45cm}Z^{X,Y}_{T_{i \mu_2}}
\end{pmatrix}}^{-1} \
\begin{pmatrix}
\vspace{0.1cm}\mathcal{O}_{V_i} \ \ \\
\mathcal{O}_{T_{i \mu_2}}
\end{pmatrix}, \quad (i \neq \mu_2)
\label{OGamma2} 
\eea
\be 
\mathcal{O}_{\Gamma}^Y = \ {(Z^{X,Y}_{\Gamma})}^{-1} \mathcal{O}_{\Gamma}, \qquad \Gamma = S, V_{\mu_2}, A_{i}, T_{ij}, \quad (i \neq j \neq \mu_2 \neq i),
\label{OGamma3}
\ee
where $X (Y)$ stands for the regularization (renormalization) scheme: $X = {\rm DR}, {\rm LR}, \ldots$, $Y = \MSbar, {\rm RI}', \ldots$ As we will see, in dimensional regularization there is no operator mixing and thus the mixing matrices are diagonal.

\bigskip

As is standard practice, the calculation of the renormalization factors of $\mathcal{O}_\Gamma$ stems from the evaluation of the corresponding one-particle-irreducible (1-PI) two-point amputated  Green's functions $\Lambda_\Gamma \equiv {\langle\psi_f\,{\mathcal O}_{\Gamma}\,\bar \psi_{f'} \rangle}_{\rm amp}$. According to the definitions of Eqs. (\ref{OGamma1} -- \ref{OGamma3}), the relations between the bare Green's functions and the renormalized ones are given by\footnote{In the right-hand sides of Eqs. (\ref{LambdaGamma1} -- \ref{LambdaGamma3}) it is, of course, understood that the regulators must be set to their limit values.} 

\bea
\begin{pmatrix}
\vspace{0.1cm}\Lambda_P^Y \ \ \\
\Lambda_{A_{\mu_2}}^Y
\end{pmatrix}
&=& \, {(Z^{X,Y}_{\psi_f})}^{1/2} \, {(Z^{X,Y}_{\psi_{f'}})}^{1/2}
{\begin{pmatrix}
\vspace{0.1cm}Z^{X,Y}_P \quad & Z^{X,Y}_{(P,A_{\mu_2})} \\
Z^{X,Y}_{(A_{\mu_2},P)} & \hspace{-0.45cm}Z^{X,Y}_{A_{\mu_2}}
\end{pmatrix}}^{-1} \
\begin{pmatrix}
\vspace{0.1cm}\Lambda_P^X \ \ \\
\Lambda_{A_{\mu_2}}^X
\end{pmatrix},
\label{LambdaGamma1} \\
\begin{pmatrix}
\vspace{0.1cm}\Lambda_{V_i}^Y \ \ \\
\Lambda_{T_{i \mu_2}}^Y
\end{pmatrix}
&=& \, {(Z^{X,Y}_{\psi_f})}^{1/2} \, {(Z^{X,Y}_{\psi_{f'}})}^{1/2}
{\begin{pmatrix}
\vspace{0.1cm}Z^{X,Y}_{V_i} \quad & Z^{X,Y}_{(V_i,T_{i \mu_2})} \\
Z^{X,Y}_{(T_{i \mu_2},V_i)} & \hspace{-0.45cm}Z^{X,Y}_{T_{i \mu_2}}
\end{pmatrix}}^{-1} \
\begin{pmatrix}
\vspace{0.1cm}\Lambda_{V_i}^X \ \ \\
\Lambda_{T_{i \mu_2}}^X
\end{pmatrix}, \quad (i \neq \mu_2)
\label{LambdaGamma2} 
\eea
\be 
\Lambda_{\Gamma}^Y = {(Z^{X,Y}_{\psi_f})}^{1/2} \, {(Z^{X,Y}_{\psi_{f'}})}^{1/2} {(Z^{X,Y}_{\Gamma})}^{-1} \Lambda_{\Gamma}^X, \quad \Gamma = S, V_{\mu_2}, A_{i}, T_{ij}, \ (i \neq j \neq \mu_2 \neq i),
\label{LambdaGamma3}
\ee
where $Z^{X,Y}_{\psi_f}, Z^{X,Y}_{\psi_{f'}}$ are the renormalization factors of the external quark fields of flavors $f$ and $f'$ respectively, defined through the relation,
\be 
\psi_{f(f')}^Y = {(Z^{X,Y}_{\psi_{f(f')}})}^{-1/2} \psi_{f(f')}.
\label{Zpsi}
\ee
We note that in the case of massless quarks, the flavor content does not affect the renormalization factors of fermion fields or the Green's functions of $\mathcal{O}_\Gamma$, and thus we omit the flavor index  in the sequel. We also note that for regularizations which break chiral symmetry (such as Wilson/clover fermions), an additive mass renormalization is also needed, beyond one loop; however, this is irrelevant for our one-loop calculations. The expressions of $\Lambda_\Gamma$ depend on the coupling constant $g_0$, whose renormalization factor is defined through
\be 
g^Y = \mu^{(D-4)/2} {(Z^{X,Y}_g)}^{-1} g_0,
\label{g_renormalized}
\ee
where $\mu$ is related to the $\MSbar$ renormalization scale $\bar{\mu}$ ($\bar{\mu} \equiv \mu {(4 \pi / e^{\gamma_E})}^{1/2}$, $\gamma_E$ is Euler's constant) and $D$ is the number of Euclidean spacetime dimensions (in DR: $D \equiv 4 - 2 \varepsilon$, in LR: $D = 4$). For our one-loop calculations, $Z^{X,Y}_g$ is set to 1 (tree-level value).

\bigskip

\begin{figure}[!ht] 
  \centering
  \includegraphics[width=16.5cm,clip]{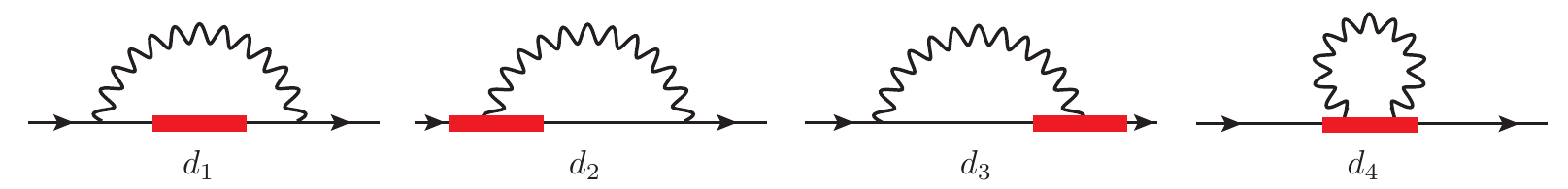} 
  \caption{Feynman diagrams contributing to the one-loop calculation of the Green's functions of staple operator $\mathcal{O}_\Gamma$. The straight (wavy) lines represent fermions (gluons). The operator insertion is denoted by a filled rectangle.} 
 \label{Fig.FeynmanDiagrams}
\end{figure}

\begin{figure}[!ht] 
  \centering
  \includegraphics[width=16.5cm,clip]{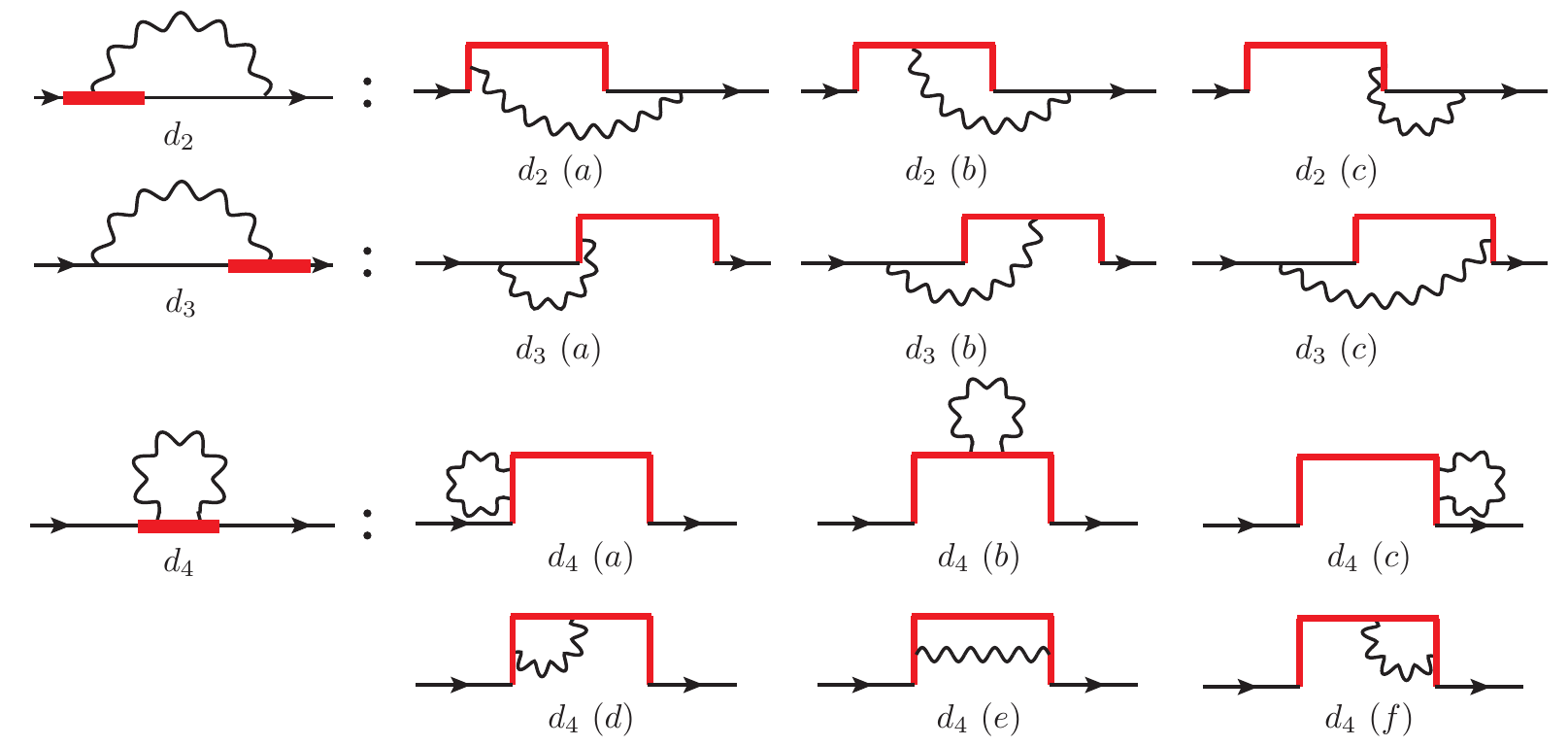}
 \caption{Subdiagrams contributing to the one-loop calculation of the Green's functions of staple operator $\mathcal{O}_\Gamma$. The straight (wavy) lines represent fermions (gluons). The operator insertion is denoted by a staple-shaped line.}
 \label{Fig.FeynmanDiagrams2}
\end{figure}

There are four one-loop Feynman diagrams contributing to $\Lambda_\Gamma$, shown in Fig. \ref{Fig.FeynmanDiagrams}. Diagrams $d_2 - d_4$ are further divided into subdiagrams, shown in Fig. \ref{Fig.FeynmanDiagrams2}, depending on the side of the staple from which gluons emanate.

\bigskip

In our computations we make use of two renormalization schemes: the modified minimal-subtraction scheme ($\MSbar$) and a variant of the modified regularization-independent scheme (RI$'$). The second one is needed for the nonperturbative evaluations of the renormalized Green's functions $\Lambda_\Gamma$ on the lattice, which will be converted to $\MSbar$, through appropriate conversion factors. For our perturbative lattice calculations, the renormalization factors of $\mathcal{O}_\Gamma$ in the $\MSbar$ scheme can be derived by calculating Eqs. (\ref{LambdaGamma1} -- \ref{LambdaGamma3}) for both $X={\rm LR}$ and $X={\rm DR}$, and demanding that their left-hand sides are $X$-independent and, thus, identical in the two regularizations.

\bigskip

For the RI$'$ scheme, we extend the standard renormalization conditions for the bilinear operators, consistently with the definitions of Eqs. (\ref{LambdaGamma1} -- \ref{LambdaGamma3}), 
\bea
{\rm tr}\Big[ 
\begin{pmatrix}
\vspace{0.1cm}\Lambda_P^{\rm{RI}'} \ \ \\
\Lambda_{A_{\mu_2}}^{\rm{RI}'}
\end{pmatrix} \ 
\begin{pmatrix}
\vspace{0.1cm}{(\Lambda_P^{\rm{tree}})}^\dagger \ \ 
{(\Lambda_{A_{\mu_2}}^{\rm{tree}})}^\dagger
\end{pmatrix}
\Big] \Bigg|_{
\begin{smallmatrix}
q_\nu = \bar{q}_\nu \\
(\forall \nu)
\end{smallmatrix}
} \! \! \! \! \! &=& \, 
{\rm tr}\Big[ 
\begin{pmatrix}
\vspace{0.1cm}\Lambda_P^{\rm{tree}} \ \ \\
\Lambda_{A_{\mu_2}}^{\rm{tree}}
\end{pmatrix} \ 
\begin{pmatrix}
\vspace{0.1cm}{(\Lambda_P^{\rm{tree}})}^\dagger \ \ 
{(\Lambda_{A_{\mu_2}}^{\rm{tree}})}^\dagger
\end{pmatrix}
\Big] = \, 4 N_c \openone_{2 \times 2}, \nonumber \\
\label{RC1} \\
{\rm tr}\Big[ 
\begin{pmatrix}
\vspace{0.1cm}\Lambda_{V_i}^{\rm{RI}'} \ \ \\
\Lambda_{T_{i \mu_2}}^{\rm{RI}'}
\end{pmatrix} \ 
\begin{pmatrix}
\vspace{0.1cm}{(\Lambda_{V_i}^{\rm{tree}})}^\dagger \ \ 
{(\Lambda_{T_{i \mu_2}}^{\rm{tree}})}^\dagger
\end{pmatrix}
\Big] \Bigg|_{
\begin{smallmatrix}
q_\nu = \bar{q}_\nu \\
(\forall \nu)
\end{smallmatrix}
} \! \! \! \! \! &=& \, 
{\rm tr}\Big[ 
\begin{pmatrix}
\vspace{0.1cm}\Lambda_{V_i}^{\rm{tree}} \ \ \\
\Lambda_{T_{i \mu_2}}^{\rm{tree}}
\end{pmatrix} \ 
\begin{pmatrix}
\vspace{0.1cm}{(\Lambda_{V_i}^{\rm{tree}})}^\dagger \ \ 
{(\Lambda_{T_{i \mu_2}}^{\rm{tree}})}^\dagger
\end{pmatrix}
\Big] = \, 4 N_c \openone_{2 \times 2}, \nonumber \\
 && \qquad \qquad \qquad \qquad \qquad \qquad \qquad (i \neq \mu_2), 
\label{RC2} 
\eea
\be
{\rm tr}\Big[ \Lambda_{\Gamma}^{\rm{RI}'}\, {(\Lambda_{\Gamma}^{\rm{tree}})}^\dagger \Big] \Bigg|_{
\begin{smallmatrix}
q_\nu = \bar{q}_\nu \\
(\forall \nu)
\end{smallmatrix}
} \! \! \! \! \! = \, 
{\rm tr}\Big[ \Lambda_{\Gamma}^{\rm{tree}}\, {(\Lambda_{\Gamma}^{\rm{tree}})}^\dagger \Big] = \, 4 N_c, \quad \Gamma = S, V_{\mu_2}, A_{i}, T_{ij}, \ (i \neq j \neq \mu_2 \neq i),
\label{RC3}
\ee
where $\Lambda_\Gamma^{\rm{tree}} \equiv \Gamma \exp (i q_{\mu_1} z)$ is the tree-level value of the Green's functions of $\mathcal{O}_\Gamma$, $\bar{q}$ is the RI$'$ renormalization scale 4-vector, and $N_c$ is the number of colors. Note that the traces appearing in Eqs. (\ref{RC1} -- \ref{RC3}) regard only Dirac and color indices; in particular, Eqs. \eqref{RC1} and \eqref{RC2} retain their $2 \times 2$ matrix form, and thus
they each correspond to four conditions. We mention that an alternative definition of the RI$'$ scheme can be adopted so that the renormalization factors depend only on a minimal set of parameters, ($\bar{q}^2, \bar{q}_{\mu_1}, \bar{q}_{\mu_2}$), rather than all the individual components of $\bar{q}$; this can be achieved by taking the average over all allowed values of the indices $i, j$, in conditions  \eqref{RC2} and \eqref{RC3}, whenever $i, j$ are present. This alternative scheme is not so useful in lattice simulations, where, besides the two special directions of the plane in which the staple lies, the temporal direction stands out from the remaining spatial directions; this leaves us with only one nonspecial direction, and thus this choice of normalization is not particularly advantageous in this case.    

\bigskip

The RI$'$ renormalization factors of fermion fields can be derived by imposing the massless normalization condition,
\be 
\textrm{tr} \Big[ S^{\rm{RI}'} {(S^{\rm{tree}})}^{-1} \Big] \Bigg|_{
\begin{smallmatrix}
q^2 = \bar{q}^2
\end{smallmatrix}
} \! \! \! \! \! = \,
\textrm{tr} \Big[ S^{\rm{tree}} {(S^{\rm{tree}})}^{-1} \Big] = \, 4 N_c,
\ee
where $S^{\rm{RI}'} \equiv \langle\psi^{\rm{RI}'} \bar{\psi}^{\rm{RI}'}\rangle$ is the RI$'$-renormalized quark propagator and $S^{\rm{tree}} \equiv {(i \slashed{q})}^{-1}$ is its tree-level value. 

\subsection{Conversion to the $\MSbar$ scheme}
The conversion of the nonperturbative RI$'$-renormalized Green's functions $\Lambda_\Gamma^{\rm{RI}'}$ to the $\MSbar$ scheme can be performed only perturbatively, since the definition of $\MSbar$ is perturbative in nature. The corresponding one-loop conversion factors between the two schemes are extracted from our calculations, and their explicit expressions are presented in Sec. \ref{Calculationprocedure}. As a consequence of the observed operator-pair mixing, some of the conversion factors will be $2 \times 2$ matrices, just as the renormalization factors of the operators. Following the definitions of Eqs. (\ref{OGamma1} -- \ref{OGamma2}), they are defined as
\bea
\begin{pmatrix} 
\vspace{0.1cm}C_P^{\MSbar, \rm{RI}'} & C_{(P, A_{\mu_2})}^{\MSbar, \rm{RI}'} \hspace{0.2cm} \\ 
C_{(A_{\mu_2},P)}^{\MSbar, \rm{RI}'} & C_{A_{\mu_2}}^{\MSbar, \rm{RI}'} \hspace{0.2cm}
\end{pmatrix} 
\! \! \! &=& 
{\begin{pmatrix} 
\vspace{0.1cm}Z^{X,\MSbar}_P \quad & Z^{X,\MSbar}_{(P,A_{\mu_2})} \\
Z^{X,\MSbar}_{(A_{\mu_2},P)} & \hspace{-0.3cm}Z^{X,\MSbar}_{A_{\mu_2}}
\end{pmatrix}}^{-1} \! \! \
{\begin{pmatrix}
\vspace{0.1cm}Z^{X, \rm{RI}'}_P \quad & Z^{X, \rm{RI}'}_{(P,A_{\mu_2})} \\
Z^{X, \rm{RI}'}_{(A_{\mu_2},P)} & \hspace{-0.3cm}Z^{X, \rm{RI}'}_{A_{\mu_2}}
\end{pmatrix}},
\label{CGamma1} \\
\begin{pmatrix}
\vspace{0.1cm}C_{V_i}^{\MSbar, \rm{RI}'} & C_{({V_i}, {T_{i \mu_2}})}^{\MSbar, \rm{RI}'} \\
C_{({T_{i \mu_2}},{V_i})}^{\MSbar, \rm{RI}'} & C_{{T_{i \mu_2}}}^{\MSbar, \rm{RI}'}
\end{pmatrix}
\! \! \! &=& 
{\begin{pmatrix}
\vspace{0.1cm}Z^{X,\MSbar}_{V_i} \quad & Z^{X,\MSbar}_{(V_i,T_{i \mu_2})} \\
Z^{X,\MSbar}_{(T_{i \mu_2},V_i)} & \hspace{-0.45cm}Z^{X,\MSbar}_{T_{i \mu_2}}
\end{pmatrix}}^{-1} \! \! \
{\begin{pmatrix}
\vspace{0.1cm}Z^{X,\rm{RI}'}_{V_i} \quad & Z^{X,\rm{RI}'}_{(V_i,T_{i \mu_2})} \\
Z^{X,\rm{RI}'}_{(T_{i \mu_2},V_i)} & \hspace{-0.45cm}Z^{X,\rm{RI}'}_{T_{i \mu_2}}
\end{pmatrix}}, (i \neq \mu_2),
\label{CGamma2} 
\eea
\be
C_{\Gamma}^{\MSbar, \rm{RI}'} = {(Z^{X,\MSbar}_{\Gamma})}^{-1} {(Z^{X,\rm{RI}'}_{\Gamma})}, \quad \Gamma = S, V_{\mu_2}, A_{i}, T_{ij}, \ (i \neq j \neq \mu_2 \neq i).
\label{CGamma3}
\ee
Being regularization independent, they can be evaluated more easily in $X = {\rm DR}$; in this regularization there is no operator mixing, and thus the conversion factors of $\mathcal{O}_\Gamma$ turn out to be diagonal. We note in passing that the definition of the $\MSbar$ scheme depends on the prescription used for extending $\gamma_5$ to D dimensions\footnote{See, e.g., Refs. \cite{Buras:1989xd,Patel:1992vu,Larin:1993tp,Larin:1993tq,Skouroupathis:2008mf,Constantinou:2013pba} for a discussion of four relevant prescriptions and some conversion factors among them.}; this, in particular, will affect conversion factors for the pseudoscalar and axial-vector operators. However, such a dependence will only appear beyond one loop. 

\bigskip

Given that the conversion factors are diagonal, the Green's functions of $\mathcal{O}_\Gamma$ in the RI$'$ scheme can be directly converted to the $\MSbar$ scheme through the following relation, valid for all $\Gamma$:
\be 
\Lambda_\Gamma^\MSbar = {(C_\psi^{\MSbar, \rm{RI}'})}^{-1} C_\Gamma^{\MSbar, \rm{RI}'} \Lambda_\Gamma^{\rm{RI}'},
\ee   
where $\mathcal{C}_{\psi}^{\MSbar,{\rm RI}'} \equiv {(Z_{\psi}^{X,\MSbar})}^{-1} Z_{\psi}^{X,{\rm RI}'}$ is the conversion factor for fermion fields.

\section{Calculation procedure and Results}
\label{Calculationprocedure}

In this section we proceed with the one-loop calculation of the renormalization factors of the staple operators in the RI$'$ and $\MSbar$ renormalization schemes, both in dimensional and lattice regularizations. We apply the prescription described above, and we present our final results. We also include the one-loop expressions for the conversion factors between the two schemes. 
 
\subsection{Calculation in dimensional regularization} 

\subsubsection{\textbf{Methodology}}
We calculate the bare Green's functions of the staple operators in $D$ Euclidean spacetime dimensions (where $D \equiv 4 - 2 \varepsilon$ and $\varepsilon$ is the regulator), in which momentum-loop integrals are well-defined. The methodology for calculating these integrals is briefly described in our previous work regarding straight Wilson-line operators \cite{Constantinou:2013pba,Spanoudes:2018zya}, and it is summarized below:
We follow the standard procedure of introducing Feynman parameters. The momentum-loop integrals depend on exponential functions of the $\mu_1$- and/or $\mu_2$-component of the internal momentum [e.g., $\exp (i p_{\mu_1} z)$, $\exp (i p_{\mu_1} \zeta)$]. The integration over the components of momentum without an exponential dependence is performed using standard D-dimensional formulae (e.g., \cite{tHooft:1973wag}), followed by a subsequent nontrivial integration over the remaining components $p_{\mu_1}$ and/or $p_{\mu_2}$. The resulting expressions contain a number of Feynman parameter integrals and/or integrals over $\zeta$-variables stemming from the definition of $\mathcal{O}_\Gamma$, which depend on modified Bessel functions of the second kind, $K_n$ and which do not have a closed analytic form; they are listed in Appendix \ref{ap.A}. We expand these expressions as Laurent series in $\varepsilon$ and we keep only terms up to $\mathcal{O} (\varepsilon^0)$. The full expressions of the bare Green's functions of $\mathcal{O}_\Gamma$ are given in Appendix \ref{ap.A}; these can be used for applying any renormalization scheme. 

\subsubsection{\textbf{Renormalization factors}}

Our one-loop results for the renormalization factors of the staple operators in both $\MSbar$ and RI$'$ schemes are presented below. 

\bigskip

In the $\MSbar$ scheme, only the pole parts [$\mathcal{O} (1/\varepsilon)$ terms] contribute to the renormalization factors. Diagram $d_1$ has no $1/\varepsilon$ terms, as it is finite in $D=4$ dimensions. Also, it gives the same expressions with the corresponding straight-line operators, because it involves only the zero-gluon operator vertex. This statement is true in any regularization. As we expected, the divergent terms arise from the remaining diagrams $d_2 - d_4$, in which end point [Eqs. (\ref{endpointdiv1}, \ref{endpointdiv2})] and cusp divergences [Eq. \eqref{cuspdiv}] arise. We provide below the pole parts for each subdiagram:
\begin{align} 
&\Lambda_\Gamma^{d_1} \vert_{1/\varepsilon} = \Lambda_\Gamma^{d_2 (a)} \vert_{1/\varepsilon} = \Lambda_\Gamma^{d_2 (b)} \vert_{1/\varepsilon} = \Lambda_\Gamma^{d_3 (b)} \vert_{1/\varepsilon} = \Lambda_\Gamma^{d_3 (c)} \vert_{1/\varepsilon} = \Lambda_\Gamma^{d_4 (e)} \vert_{1/\varepsilon} = 0, \\ 
&\Lambda_\Gamma^{d_2 (c)} \vert_{1/\varepsilon} = \Lambda_\Gamma^{d_3 (a)} \vert_{1/\varepsilon} = \frac{g^2 C_F}{16 \pi^2} \Lambda_\Gamma^{\rm tree} \frac{1}{\varepsilon} (1 - \beta), \label{endpointdiv1} \\ 
&\Lambda_\Gamma^{d_4 (a)} \vert_{1/\varepsilon} = \Lambda_\Gamma^{d_4 (b)} \vert_{1/\varepsilon} = \Lambda_\Gamma^{d_4 (c)} \vert_{1/\varepsilon} = \frac{g^2 C_F}{16 \pi^2} \Lambda_\Gamma^{\rm tree} \frac{1}{\varepsilon} (2 + \beta), \label{endpointdiv2} \\
&\Lambda_\Gamma^{d_4 (d)} \vert_{1/\varepsilon} = \Lambda_\Gamma^{d_4 (f)} \vert_{1/\varepsilon} = \frac{g^2 C_F}{16 \pi^2} \Lambda_\Gamma^{\rm tree} \frac{1}{\varepsilon} (- \beta), \label{cuspdiv}
\end{align}
where $C_F = (N_c^2 - 1)/ (2 N_c)$ and $\beta$ is the gauge fixing parameter , defined such that $\beta = 0 \ (1)$ corresponds to the Feynman (Landau) gauge. It is deduced that diagrams $d_2, d_3$ give the same pole terms as in the case $y=0$, since only end points affect these diagrams (no cusps). Also, the result for the cusp divergences of angle $\pi/2$ [Eq. \eqref{cuspdiv}] agree with previous studies of nonsmooth Wilson-line operators for a general cusp angle $\theta$ \cite{Brandt:1981kf, Knauss:1984rx, Korchemsky:1987wg}: it follows from these studies that the one-loop result corresponding to each of the diagrams $d_4 (d)$ and $d_4 (f)$ is given by $- (g^2 C_F)/(16 \pi^2 \varepsilon) \ (2 \theta \cot \theta + \beta)$. By imposing that the $\MSbar$-renormalized Green's functions of $\mathcal{O}_\Gamma$ are equal to the finite parts (exclude pole terms) of the corresponding bare Green's functions, we derive the renormalization factors of $\mathcal{O}_\Gamma$ in $\MSbar$, using Eqs. (\ref{LambdaGamma1} -- \ref{LambdaGamma3}); the result is given below,
\be
Z_\Gamma^{{\rm DR},\MSbar} = 1 + \frac{g^2 C_F}{16 \pi^2} \frac{7}{\varepsilon} + \mathcal{O} (g^4),
\label{ZGammaDRMSbar}
\ee 
where we make use of the one-loop expression for the renormalization factor $Z_\psi^{\rm{DR},\MSbar}$, given in Appendix \ref{ap.B} [Eq. \eqref{ZpsiDRMSbar}]. Since the pole parts are multiples of the tree-level values $\Lambda_\Gamma^{\rm tree}$, the nondiagonal elements of the $\MSbar$ renormalization factors, defined in Eqs. (\ref{OGamma1}, \ref{OGamma2}), are equal to zero. The diagonal elements, shown in Eq. \eqref{ZGammaDRMSbar}, depend neither on the Dirac structure, nor on the lengths of the staple segments; further, they are gauge invariant.

\bigskip

In the RI$'$ scheme, there are additional finite terms, which contribute to the renormalization factors of $\mathcal{O}_\Gamma$ [according to the conditions of Eqs. (\ref{RC1} -- \ref{RC3})]. These terms depend on the external momentum, and they stem from all Feynman diagrams. They are also multiples of the tree-level values of the Green's functions. As a consequence, the RI$'$ mixing matrices, defined in Eqs. (\ref{OGamma1}, \ref{OGamma2}), are also diagonal. Therefore, there is no operator mixing in DR. The results for $Z_\Gamma^{\rm{DR, RI}'}$, together with $Z_\Gamma^{\rm{DR}, \MSbar}$ [Eq. \eqref{ZGammaDRMSbar}], lead directly to the conversion factors $C_\Gamma^{\MSbar,\rm{RI}'}$ through the relation,
\be 
Z_\Gamma^{\rm{DR,RI}'} = C_\Gamma^{\MSbar,\rm{RI}'} + \frac{g^2 C_F}{16 \pi^2} \frac{7}{\varepsilon} + \mathcal{O} (g^4).
\label{ZGammaDRRI'}
\ee 
Our resulting expressions for the conversion factors are given in the following subsection [Eqs. (\ref{C_S} -- \ref{C_rest})].

\subsubsection{\textbf{Conversion factors}}

We present below our results for the conversion factors of staple operators between the RI$'$ and $\MSbar$ schemes. Since the renormalization factors of $\mathcal{O}_\Gamma$ are diagonal in both $\MSbar$ and RI$'$ schemes, the conversion factors will also be diagonal. Our expressions depend on integrals of modified Bessel functions of the second kind $K_n$, over one Feynman parameter and possibly over one of the variables $\zeta$ appearing in Eq. \eqref{stapleW}. These integrals are denoted by $P_i \equiv P_i (\bar{q}^2, \bar{q}_{\mu_1}, z)$, $Q_i \equiv Q_i (\bar{q}^2, \bar{q}_{\mu_1}, \bar{q}_{\mu_2}, z, y)$ and $R_i \equiv R_i (\bar{q}^2, \bar{q}_{\mu_1}, \bar{q}_{\mu_2}, z, y)$; they are defined in Eqs. (\ref{P1} -- \ref{R8}) of Appendix \ref{ap.A}.
\begin{align}
C_{S}^{\text{RI}', \MSbar} = 1 + \frac{g^2 C_F}{16 \pi^2} \boldmath\Bigg\{ &(15 - \beta) + 2 (\beta +6) \gamma_E + 7 \log \left(\frac{\overline{\mu}^2}{\overline{q}^2}\right) + (\beta +2) \log \left(\frac{\overline{q}^2 z^2}{4}\right) \nonumber \\
& + 4 \log \left(\frac{\overline{q}^2 y^2}{4}\right) +4 \left(2 \ \frac{y}{z} \tan^{-1}\left(\frac{y}{z}\right)-\log\left(1+\frac{y^2}{z^2}\right)\right) + 2 (\beta +2) P_1 \nonumber \\
& -2 \beta \sqrt{\overline{q}^2} \left| z\right| P_4 -2 i \overline{q}_{\mu_1} \Big(2 Q_1-\beta  z (P_1-P_2)\Big) + 4 \overline{q}_{\mu_2} (R_6 - R_2) \boldmath\Bigg\} + \mathcal{O} (g^4)
\label{C_S}
\end{align}
\begin{align}
C_{V_{\mu_1}}^{\text{RI}', \MSbar} = 1 + \frac{g^2 C_F}{16 \pi^2} \boldmath\Bigg\{ &15 + 2 (\beta +6) \gamma_E + 7 \log \left(\frac{\overline{\mu}^2}{\overline{q}^2}\right) + (\beta +2) \log \left(\frac{\overline{q}^2 z^2}{4}\right) + 4 \log \left(\frac{\overline{q}^2 y^2}{4}\right) \nonumber \\
& +4 \left(2 \ \frac{y}{z} \tan^{-1}\left(\frac{y}{z}\right)-\log\left(1+\frac{y^2}{z^2}\right)\right) + 2 (\beta  P_1-2 P_2) \nonumber \\
&-2 \sqrt{\overline{q}^2} \left| z\right| (\beta  P_4-2 P_5)-\frac{1}{2} \beta \overline{q}^2 z^2 (P_1-P_2) +4 \overline{q}_{\mu_2} (R_6-R_2) \nonumber \\
& -i \overline{q}_{\mu_1} \bigg(4 Q_1 - 2 z \Big(\beta (P_1-P_2)-4 P_3\Big) + \beta \sqrt{\overline{q}^2} \ z \left| z\right| (P_4-P_5)\bigg) \nonumber \\
&+ \overline{q}_{\mu_1}^2 \bigg(2 \frac{\left| z\right|}{\sqrt{q^2}} \Big((\beta -2) P_4+2 P_5\Big) + \beta z^2 P_3 \bigg) \boldmath\Bigg\} + \mathcal{O} (g^4)
\label{C_V1}
\end{align}
\begin{align}
C_{V_{\nu}}^{\text{RI}', \MSbar} = 1 + \frac{g^2 C_F}{16 \pi^2} \boldmath\Bigg\{ &(15 - \beta) + 2 (\beta +6) \gamma_E + 7 \log \left(\frac{\overline{\mu}^2}{\overline{q}^2}\right) + (\beta +2) \log \left(\frac{\overline{q}^2 z^2}{4}\right) \nonumber \\
& + 4 \log \left(\frac{\overline{q}^2 y^2}{4}\right) +4 \left(2 \ \frac{y}{z} \tan^{-1}\left(\frac{y}{z}\right)-\log\left(1+\frac{y^2}{z^2}\right)\right) + 2 (\beta  P_1-2 P_2) \nonumber \\
& - \beta \sqrt{\overline{q}^2} \left| z\right| P_4 -2 i \overline{q}_{\mu_1} \Big(2 Q_1-\beta  z (P_1-P_2)\Big) + 4 \overline{q}_{\mu_2} (R_6 - R_2) \nonumber \\
&+ \overline{q}_{\nu}^2 \bigg(2 \frac{\left| z\right|}{\sqrt{q^2}} \Big((\beta -2) P_4+2 P_5\Big) - \beta z^2 P_3 \bigg) \boldmath\Bigg\} + \mathcal{O} (g^4), \quad (\nu \neq \mu_1) 
\label{C_V2}
\end{align}
\begin{align}
C_{T_{\mu_1 \nu}}^{\text{RI}', \MSbar} = 1 + \frac{g^2 C_F}{16 \pi^2} \boldmath\Bigg\{ &15 + 2 (\beta +6) \gamma_E + 7 \log \left(\frac{\overline{\mu}^2}{\overline{q}^2}\right) + (\beta +2) \log \left(\frac{\overline{q}^2 z^2}{4}\right) + 4 \log \left(\frac{\overline{q}^2 y^2}{4}\right) \nonumber \\
& +4 \left(2 \ \frac{y}{z} \tan^{-1}\left(\frac{y}{z}\right)-\log\left(1+\frac{y^2}{z^2}\right)\right) + 2 (\beta -2) P_1 - \beta \sqrt{\overline{q}^2} \left| z\right| P_4 \nonumber \\
&-\frac{1}{2} \beta \overline{q}^2 z^2 (P_1-P_2) + 4 \overline{q}_{\mu_2} (R_6 - R_2) \nonumber \\
&-i \overline{q}_{\mu_1} \Big(4 Q_1 - 2 \beta z (P_1-P_2) + \beta \sqrt{\overline{q}^2} \ z \left| z\right| (P_4-P_5)\Big) \nonumber \\
&+\beta z^2 \left(\overline{q}_{\mu_1}^2+\overline{q}_{\nu}^2\right) P_3  \boldmath\Bigg\} + \mathcal{O} (g^4), \quad (\nu \neq \mu_1)
\label{C_T}
\end{align}
\be 
C_P = C_S, \quad C_{A_{\mu_1}} = C_{V_{\mu_1}}, \quad C_{A_\nu} = C_{V_\nu}, \quad C_{T_{\rho \sigma}} = C_{T_{\mu_1 \nu}}, \quad (\mu_1, \nu, \rho, \sigma \ \text{are all different}).
\label{C_rest}
\ee
We note that the real parts of the above expressions, as well as the bare Green's functions, are not analytic functions of z (y) near $z \rightarrow 0$ ($y \rightarrow 0$); in particular, the limit $z \rightarrow 0$ leads to quadratic divergences, while the limit $y \rightarrow 0$  leads to logarithmic divergences. The singular limits were expected, due to the appearance of contact terms beyond tree level. In the case $y = 0$, the staple operators are replaced by straight-line operators of length $|z|$, the renormalization of which is addressed in our work of Ref. \cite{Constantinou:2017sej}. In the case $z = 0$, the nonlocal operators are replaced by local bilinear operators, the renormalization of which is studied, e.g., in Refs. \cite{Larin:1993tq,Gracey:2003yr,Skouroupathis:2007jd,Skouroupathis:2008mf}.

\bigskip

\begin{figure} [!ht]
\centering
\includegraphics[width=8.1cm]{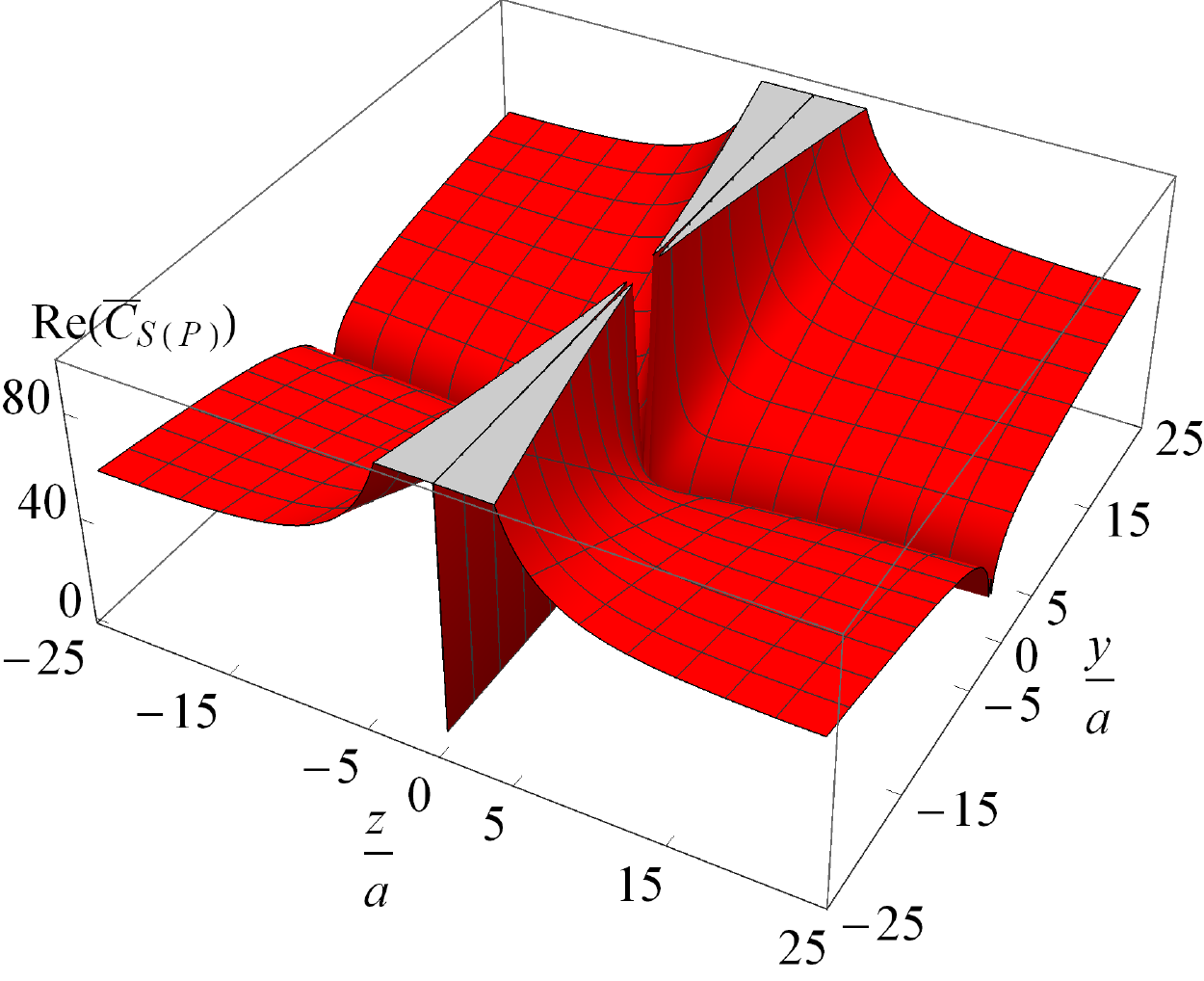} \hspace{0.1cm}
\includegraphics[width=7.9cm]{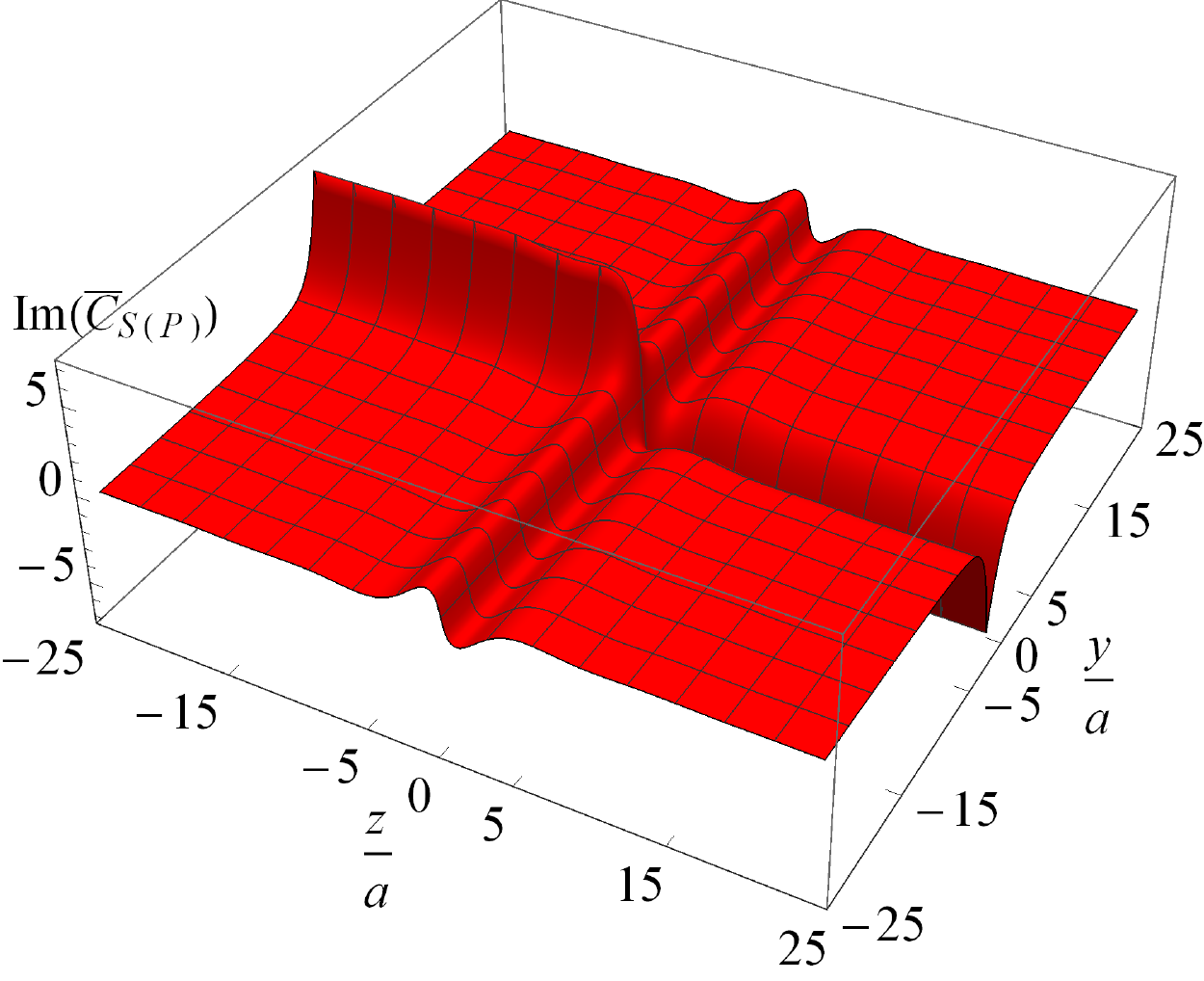} \\
\includegraphics[width=8.1cm]{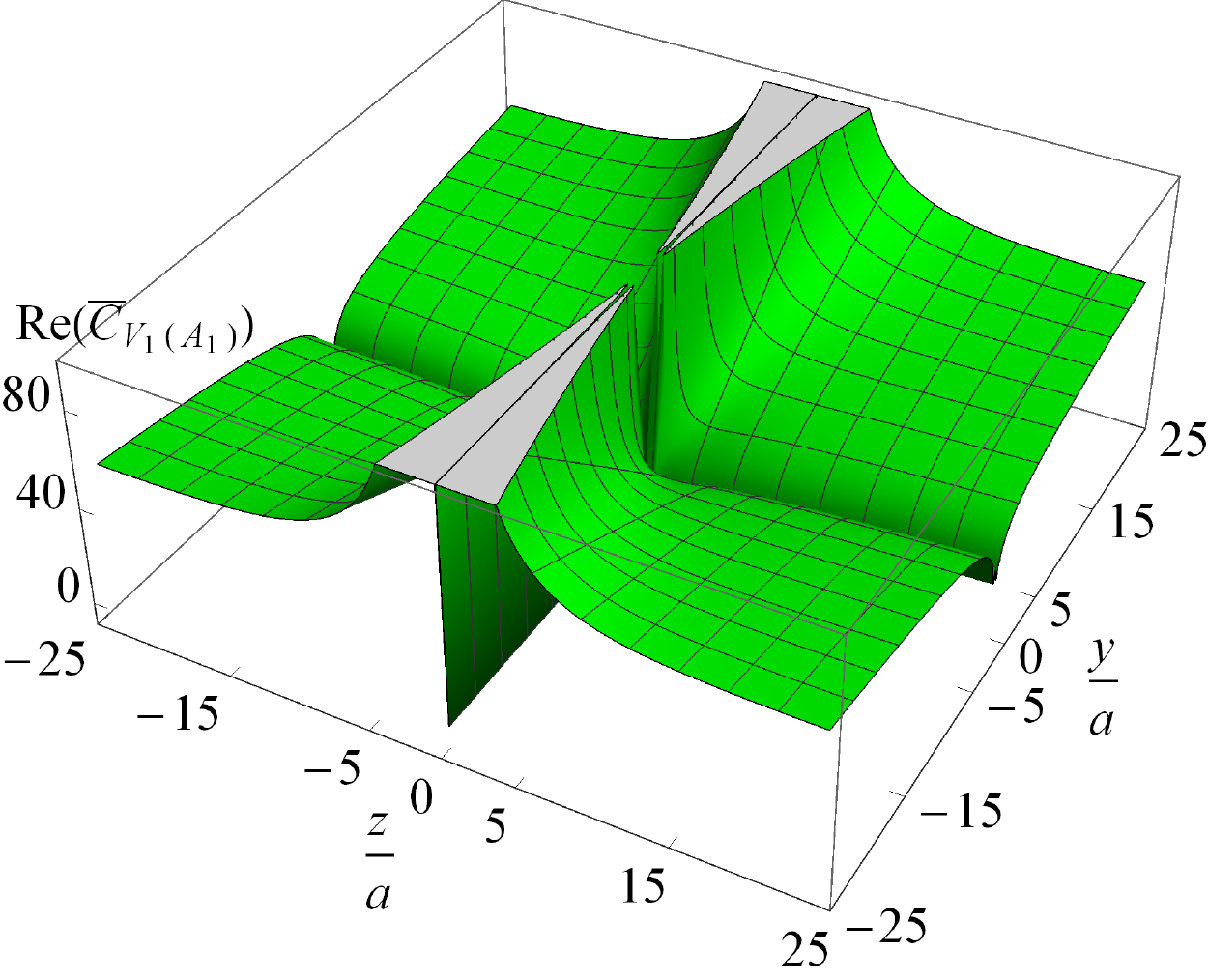} \hspace{0.1cm}
\includegraphics[width=7.9cm]{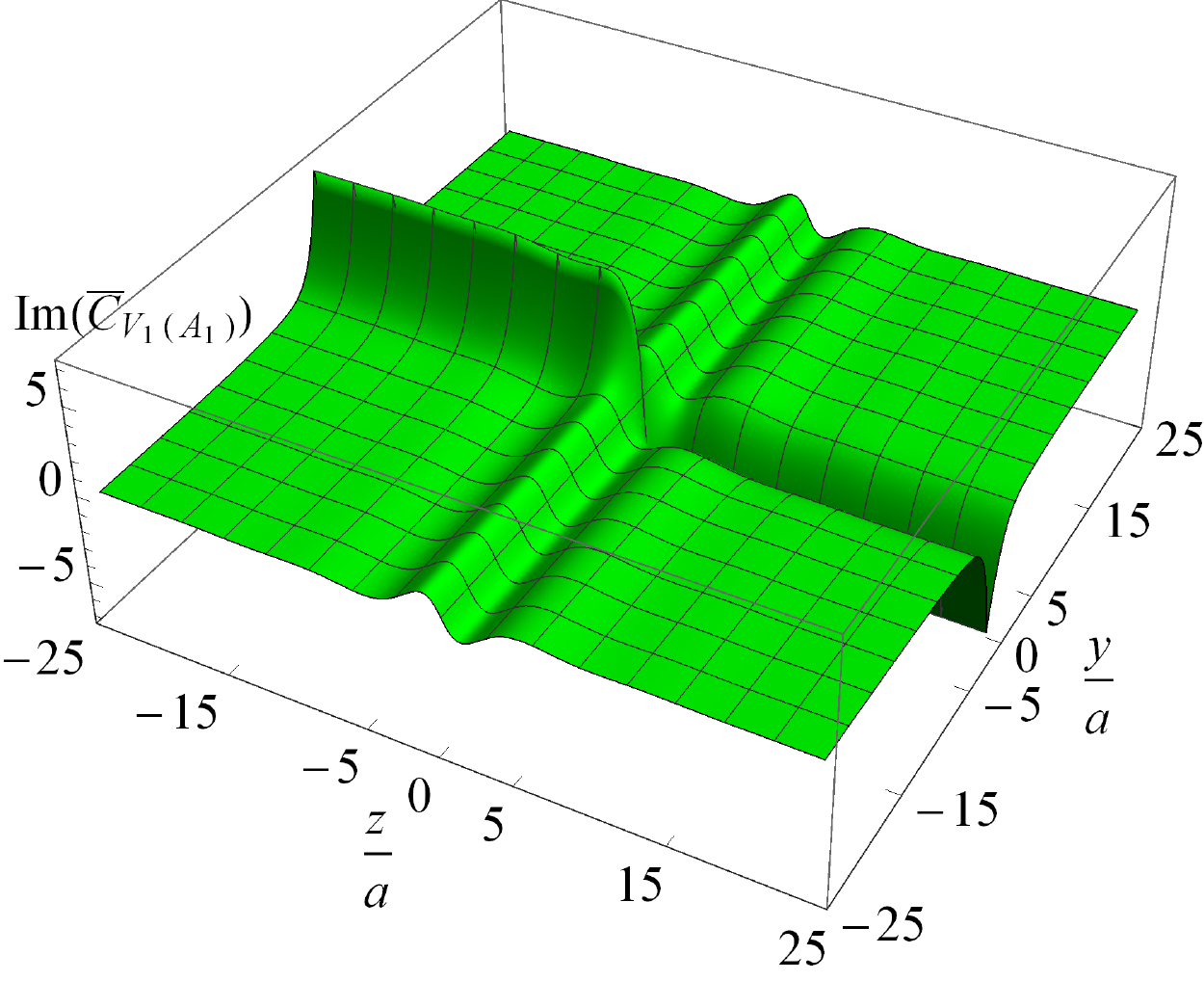} \\
\includegraphics[width=7.8cm]{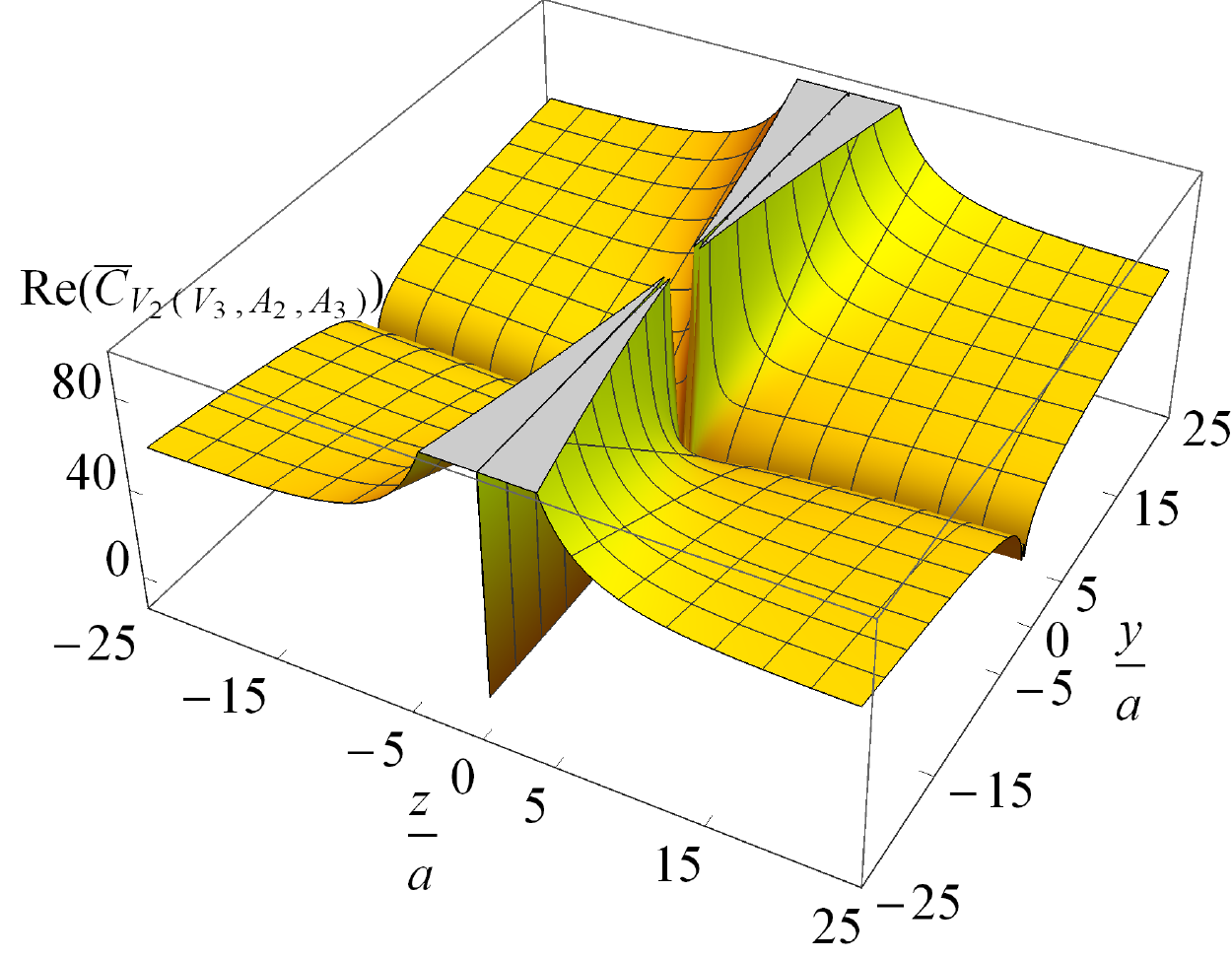} \hspace{0.1cm}
\includegraphics[width=8.2cm]{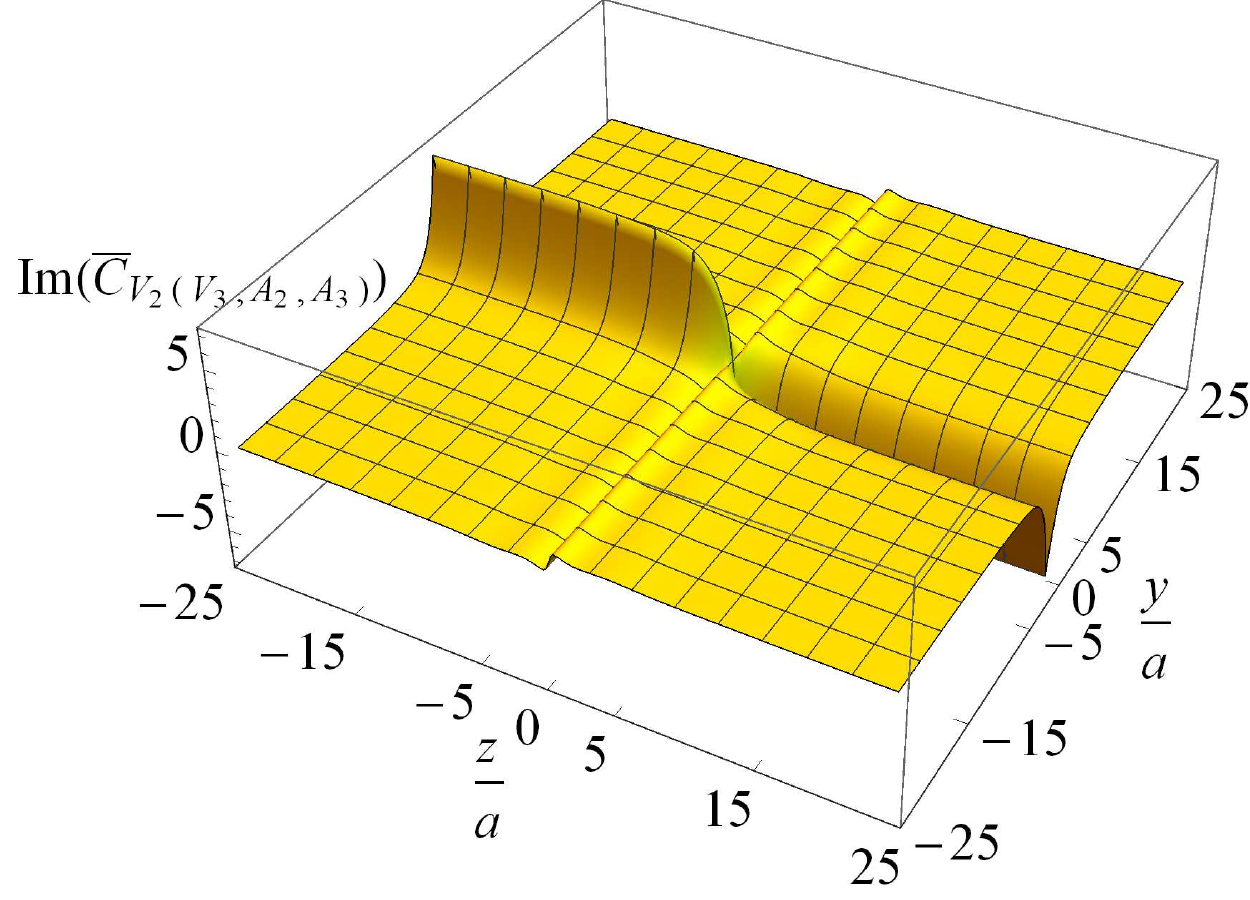} 
\caption{Real (left panels) and imaginary (right panels) parts of the quantities $\overline{C}_{S} = \overline{C}_{P}$, $\overline{C}_{V_1} = \overline{C}_{A_1}$ and $\overline{C}_{V_2} = \overline{C}_{V_3} = \overline{C}_{A_2} = \overline{C}_{A_3}$, involved in the one-loop expressions of the conversion factors: $C^{{\rm RI}', \MSbar}_\Gamma = 1 + \frac{g^2 C_F}{16 \pi^2} \ \overline{C}_\Gamma + \mathcal{O} (g^4)$, as functions of $z/a$ and $y/a$ [for $\beta = 1$, $\bar{\mu} = 2 \ GeV$, $a = 0.09$ \ fm, $a \bar{q} = (\frac{2 \pi}{L} n_1, \frac{2 \pi}{L} n_2, \frac{2 \pi}{L} n_3, \frac{2 \pi}{T} (n_4 + \frac{1}{2}))$, $L=32$, $T=64$, $(n_1, n_2, n_3, n_4) = (4,4,4,9)$]. Here, we choose $\mu_1 = 1$ and $\mu_2 = 2$.}
\label{plot}
\end{figure}

Since our results for the conversion factors will be combined with nonperturbative data, it is useful to employ certain values of the free parameters mostly used in simulations. To this end, we set: $\bar{\mu} = 2$ GeV and $\beta = 1$ (Landau gauge). For the RI$'$ scale we employ values which are relevant for simulations by ETMC \cite{Alexandrou:2016jqi}, as follows: $a \bar{q} = (\frac{2 \pi}{L} n_1, \frac{2 \pi}{L} n_2, \frac{2 \pi}{L} n_3, \frac{2 \pi}{T} (n_4 + \frac{1}{2}))$, where $a$ is the lattice spacing, ($L^3 \times T$) is the lattice size and $(n_1, n_2, n_3, n_4)$ is a 4-vector defined on the lattice. A standard choice of values for $n_i$ is the case $n_1 = n_2 = n_3 \neq n_4$, in which the temporal component $n_4$ stands out from the remaining equal spatial components. As an example we apply $(n_1, n_2, n_3, n_4) = (4,4,4,9)$, $L=32$, $T=64$ and $a = 0.09$ \ fm. For a better assessment of our results, we plot in Fig. \ref{plot} the real and imaginary parts of the quantities $\overline{C}_\Gamma$, defined through $C^{RI', \MSbar}_\Gamma = 1 + \frac{g^2 C_F}{16 \pi^2} \ \overline{C}_\Gamma + \mathcal{O} (g^4)$, as functions of the dimensionless variables $z/a$ and $y/a$, using the above parameter values. In the case $y = 0$, we use the expressions of the conversion factors for straight-line operators, calculated in Ref. \cite{Constantinou:2017sej}, while in the case $z = 0$, we use the one-loop expressions of the conversion factors for local bilinear operators, written in Refs. \cite{Skouroupathis:2007jd,Skouroupathis:2008mf}. For definiteness, we choose $\mu_1 = 1$ and $\mu_2 = 2$. Graphs for $\overline{C}_{V_4} = \overline{C}_{A_4}$, $\overline{C}_{T_{12}} = \overline{C}_{T_{13}} = \overline{C}_{T_{42}} = \overline{C}_{T_{34}}$ and $\overline{C}_{T_{14}} = \overline{C}_{T_{32}}$ are not included in Fig. \ref{plot}, as their resulting values are very close to those of $\overline{C}_{V_2}$ (fractional differences: $\lesssim 10^{-3}$).       

\bigskip

The real parts of $\overline{C}_\Gamma$ are even functions of both $z/a$ and $y/a$. In Fig. \ref{plot}, one observes that, for large values of $z/a$, they tend to stabilize, while for large values of $y/a$ they tend to increase; thus, a two-loop calculation of the conversion factors is essential for more sufficiently convergent results. Further, the dependence on the choice of $\Gamma$ becomes milder for increasing values of $z/a$ and $y/a$. Regarding the imaginary parts of $\overline{C}_\Gamma$, they are odd functions of $z/a$ and even functions of $y/a$. For large values of $z/a$ or $y/a$, they tend to converge to a positive value. In particular, when both $z/a$ and $y/a$ take large values, the imaginary parts tend to zero. For large values of $y/a$ and, simultaneously, small values of $z/a$, the imaginary parts of $\overline{C}_\Gamma$ demonstrate a small fluctuation around zero, which differs for each $\Gamma$, either in form (e.g., $\overline{C}_{V_2}$ and $\overline{C}_{S}$ have opposite signs for given values of $z/a$, $y/a$) or in magnitude (e.g., the fluctuation of $\overline{C}_{S}$ is bigger and sharper than the fluctuation of $\overline{C}_{V_1}$). As regards the $\bar{q}$ dependence, we have not included further graphs for the sake of conciseness; however, testing a variety of values for the components of $a \bar{q}$, used in simulations, we find no significant difference, especially for large values of $z/a$ and $y/a$.

\subsection{Calculation in lattice regularization}

\subsubsection{\textbf{Methodology}}
At first, let us give the lattice version of the staple operators,
\begin{gather}
\mathcal{O}_\Gamma^{\rm latt.} \equiv \bar\psi(x) \ \Gamma \ W(x,x+m a \hat{\mu}_2,x+m a \hat{\mu}_2+n a  \hat{\mu}_1,x+n a \hat{\mu}_1) \ \psi(x + n a {\hat{\mu}}_1),  
\end{gather}
\begin{align}
W(&x, x+m a \hat{\mu}_2, x + m a \hat{\mu}_2 + n a \hat{\mu}_1, x + n a \hat{\mu}_1) \equiv \nonumber \\
&{\Big(\prod_{\ell=0}^{m \mp 1} U_{\pm \mu_2} (x + \ell a \hat{\mu}_2)\Big)} \cdot {\Big(\prod_{\ell=0}^{n \mp 1} U_{\pm \mu_1} (x + m a \hat{\mu}_2 + \ell a \hat{\mu}_1)\Big)} \cdot {\Big(\prod_{\ell=0}^{m \mp 1} U_{\pm \mu_2} (x + n a \hat{\mu}_1 + \ell a \hat{\mu}_2)\Big)}^\dagger, \nonumber \\
& \hspace{11cm} n \equiv z/a, \ m \equiv y/a,
\end{align}
where upper (lower) signs of the first and third parenthesis correspond to $m>0$ ($m<0$) and upper (lower) signs of the second parenthesis correspond to $n>0$ ($n<0$). The calculation of the bare Green's functions of such nonlocal operators on the lattice is more complicated than the corresponding calculation of local operators; the products of gluon links lead to expressions whose summands, taken individually, contain possible additional IR singularities along a whole hyperplane, instead of a single point [terms $\sim 1/\sin(p_\mu / 2)$ or $1/\sin^2(p_\mu / 2)$]. Also, the UV-regulator limit, $a \rightarrow 0$, is more delicate in this case, as the Green's functions depend on $a$ through the additional combinations $z/a$, $y/a$, besides the combination $a q$ (where $q$ is the external quark momentum). Thus, we have to modify  the standard methods of evaluating Feynman diagrams on the lattice \cite{Kawai:1980ja}, in order to apply them in the case of nonlocal operators. 

\bigskip

The procedure that we used for the calculation of the bare Green's functions of $\mathcal{O}_\Gamma^{\rm latt.}$ is briefly described in our previous work regarding straight Wilson-line operators \cite{Constantinou:2017sej}, and it is summarized below:
The main task is to write the lattice expressions, in terms of continuum integrals, which are easier to calculate, plus lattice integrals independent of $a q$; however, the latter will still have a nontrivial dependence on $z/a$ and $y/a$. To this end, we perform a series of additions and subtractions to the original integrands: we extend the standard procedure of Kawai \textit{et al.} \cite{Kawai:1980ja}, in order to isolate the possible IR divergences stemming from the integration over the $p_{\mu}$ component, which appears on the integrals' denominators [$\sim 1/\sin(p_\mu / 2)$ or $1/\sin^2(p_\mu / 2)$]. To accomplish this, we add and subtract to the original integrands the lowest order of their Laurent expansion in $p_{\mu}$. Also, in order to end up with continuum integrals, we add and subtract the continuum counterparts of the integrands; then, the integration region can be split up into two parts: the whole domain of the real numbers minus the region outside the Brillouin zone. The above operations allow us to separate the original expressions into a sum of two parts: one part contains integrals which can be evaluated explicitly for nonzero values of $a$, leading to linear or logarithmic divergences, and a second part for which a naive $a \rightarrow 0$ limit can be taken, e.g., 
\be 
\int dp f(p) e^{i (z/a) p} \rightarrow 0, \quad \int dp f(p) \sin^2(\frac{z}{a} p) \rightarrow \frac{1}{2}\int dp f(p).
\ee
The numerical integrations entail a very small systematic error, which is smaller than the last digit presented in all results shown in the sequel.

\subsubsection{\textbf{Green's functions and operator mixing}}
\label{sec.Mixing}

The results for the bare lattice Green's functions of the staple operators are presented below in terms of the $\MSbar$-renormalized Green's functions, derived by the corresponding calculation in DR, 
\be
\Lambda_\Gamma^{\rm LR} = \Lambda_\Gamma^{\MSbar} - \frac{g^2\,C_F}{16\,\pi^2}\, e^{i\,q_{\mu_1} z}\, \cdot \mathcal{F} \ + \ \mathcal{O} (g^4),
\label{Lambda_LR1}
\ee
\be
\mathcal{F} = \Big[\Gamma \Big(\alpha_1 + 3.7920 \, \beta + \alpha_2\,\frac{|z| + 2 \ |y|}{a} + \log \left(a^2 \bar\mu^2\right) \left(8-\beta\right) \Big) + \text{sgn}(y) \Big[ \Gamma, \gamma_{\mu_2} \Big] \,\Big(\alpha_3 + \alpha_4\,c_{\rm SW}\Big) \Big],
\label{Lambda_LR2}
\ee
where $\alpha_i$ are numerical constants which depend on the gluon action in use; their values are given in Table \ref{tab:stapleLR} for the Wilson, Tree-level Symanzik and Iwasaki gluon actions\footnote{A more precise result for the numerical constant $3.7920$, which multiplies the $\beta$ parameter, is $16 \pi^2 P_2$, where $P_2 = 0.02401318111946489(1)$ \cite{Luscher:1995np}.}. We note that $\alpha_2$, $\alpha_3$, and $\alpha_4$ have the same values (up to a sign) as the corresponding coefficients in the straight-line operators \cite{Constantinou:2017sej}. 
\begin{table}[thb]
  \centering
  \begin{tabular}{|l|l|l|l|l|}
  \hline
\ \textbf{Gluon action} & \ \quad \ $\boldsymbol{\alpha_1}$ & \ \quad \ $\boldsymbol{\alpha_2}$ & \ \quad \ $\boldsymbol{\alpha_3}$ & \ \quad \ $\boldsymbol{\alpha_4}$  \\
\hline
\hline
\ Wilson & \ -22.5054 \ & \ \ 19.9548 \ & \ \ \ 7.2250 \ & \ \ -4.1423 \ \\
\ Tree-level Symanzik \ & \ -22.0931 \ & \ \ 17.2937 \ & \ \ \ 6.3779 \ & \ \ -3.8368 \ \\ 
\ Iwasaki & \ -18.2456 \ & \ \ 12.9781 \ & \ \ \ 4.9683 \ & \ \ -3.2638 \ \\
\hline
  \end{tabular}
  \caption{Numerical values of the coefficients $\alpha_1{-}\alpha_4$ appearing in the one-loop bare lattice Green's functions $\Lambda_\Gamma^{\rm LR}$.}
  \label{tab:stapleLR}
\end{table}

In Eqs. (\ref{Lambda_LR1}, \ref{Lambda_LR2}), we observe that there is a linear divergence [$\mathcal{O} (1/a)$], which depends on the length of the staple line ($|z| + 2 \ |y|$); this was expected according to the studies of closed Wilson-loop operators in regularizations other than DR \cite{Dotsenko:1979wb}. This divergence arises from the tadpolelike diagram $d_4$ and in particular from the subdiagrams $d_4 (a)$, $d_4 (b)$, $d_4 (c)$. We note that the coefficient $\alpha_2$ entering the strength of the linear divergence, is given by 
\be 
\alpha_2 = - \frac{1}{2} \ \int_{-\pi}^\pi \frac{d^3 p}{(2 \pi)^3} \ {D(\bar{p})}_{\nu\nu}, 
\ee
where ${D(p)}_{\mu\nu}$ is the gluon propagator, $\hat{\nu}$ is the direction parallel to each straight-line segment of the Wilson line and $\bar{p}$ equals the four-vector momentum $p$ with $p_\nu \rightarrow 0$. Moreover, additional contributions of different Dirac structures than the original operators appear ($[\Gamma, \gamma_{\mu_2}]$ terms); these contributions arise from the ``sail'' diagrams $d_2$, $d_3$ and in particular from the subdiagrams $d_2 (c)$, $d_3 (a)$. In order to obtain on the lattice the same results for the $\MSbar$-renormalized Green's functions as those obtained in DR, we have to subtract such regularization dependent terms in the renormalization process. A simple multiplicative renormalization cannot eliminate these terms; the introduction of mixing matrices is therefore necessary. However, for the operators with $\Gamma = S, V_{\mu_2}, A_{i}, T_{ij}$, where $i \neq j \neq \mu_2 \neq i$, the contribution $[\Gamma, \gamma_{\mu_2}]$ is zero, and, thus, there is no mixing for these operators. In conclusion, there is mixing between the operators $(\mathcal{O}_P, \mathcal{O}_{A_{\mu_2}}), (\mathcal{O}_{V_i}, \mathcal{O}_{T_{i\mu_2}})$, where $i \neq \mu_2$, as we have mentioned previously. This feature must be taken into account in the nonperturbative renormalization of TMDs.     

\subsubsection{\textbf{Renormalization factors}}

The $\MSbar$ renormalization factors can be derived by the requirement that the terms in Eq. \eqref{Lambda_LR2} vanish in the renormalized Green's functions. Thus, through Eqs. (\ref{LambdaGamma1} -- \ref{LambdaGamma3}), one obtains the following results for the diagonal and nondiagonal elements of the renormalization factors:  
\be 
Z_{\Gamma}^{{\rm LR}, \MSbar} = 1 + \frac{g^2 C_F}{16 \pi^2} \Big[ (e_1^\psi + 1 - \alpha_1) - \alpha_2 \ \frac{|z| + 2 \ |y|}{a} + e_2^\psi c_{SW} + e_3^\psi c_{SW}^2 - 7 \log \left(a^2 \bar\mu^2\right) \Big] + \mathcal{O} (g^4),
\label{ZGammaLRMSbar}
\ee 
\be 
Z_{(P, A_{\mu_2})}^{{\rm LR}, \MSbar} = Z_{(A_{\mu_2},P)}^{{\rm LR}, \MSbar} = Z_{(V_i, T_{i \mu_2})}^{{\rm LR}, \MSbar} = Z_{(T_{i \mu_2},V_i)}^{{\rm LR}, \MSbar} = \frac{g^2 C_F}{16 \pi^2} \text{sgn} (y) (-2) \Big[ \alpha_3 + \alpha_4 c_{SW} \Big] + \mathcal{O} (g^4),
\ee
where the coefficients $e_i^\psi$ stem from the renormalization factor of the fermion field $Z_\psi^{{\rm LR}, \MSbar}$, given in Appendix \ref{ap.B} [Eq. \eqref{ZpsiLRMSbar}]. 

\bigskip

A number of observations are in order, regarding the above one-loop results: both diagonal and nondiagonal elements of the renormalization factors are operator independent, just as the corresponding renormalization factors in DR. Also, the dependence of the diagonal elements on the clover coefficient $c_{SW}$ is entirely due to the renormalization factor of fermion fields; on the contrary, the dependence of the nondiagonal elements on $c_{SW}$ is derived from the Green's functions of the operators, and in particular it is different for each choice of gluon action. Consequently, tuning the clover coefficient we can set the nondiagonal elements of the renormalization factors to zero and, thus, suppress the operator mixing. At one-loop level, this can be done by choosing $c_{SW} = - \alpha_3/ \alpha_4$. For the gluon actions given in this paper, the values of the coefficient $c_{SW}$, which lead to no mixing at one loop, are 1.7442 for Wilson action, 1.6623 for tree-level Symanzik action and 1.5222 for Iwasaki action; these values are the same as those, which eliminate the mixing in the case of straight-line operators \cite{Constantinou:2017sej}. 

\bigskip

In the RI$'$ scheme, the renormalization factors can be read off our expressions for the conversion factors, given in Eqs. (\ref{C_S} -- \ref{C_rest}), in a rather straightforward way,
\be 
Z_\Gamma^{\rm{LR,RI}'} = C_\Gamma^{\MSbar,\rm{RI}'} + \frac{g^2 C_F}{16 \pi^2} \Big[ (e_1^\psi + 1 - \alpha_1) - \alpha_2 \ \frac{|z| + 2 \ |y|}{a} + e_2^\psi c_{SW} + e_3^\psi c_{SW}^2 - 7 \log \left(a^2 \bar\mu^2\right) \Big] + \mathcal{O} (g^4),
\ee  
\be 
Z_{(P, A_{\mu_2})}^{{\rm LR}, {\rm RI}'} = Z_{(A_{\mu_2},P)}^{{\rm LR}, {\rm RI}'} = Z_{(V_i, T_{i \mu_2})}^{{\rm LR}, {\rm RI}'} = Z_{(T_{i \mu_2},V_i)}^{{\rm LR}, {\rm RI}'} = \frac{g^2 C_F}{16 \pi^2} \text{sgn} (y) (-2) \Big[ \alpha_3 + \alpha_4 c_{SW} \Big] + \mathcal{O} (g^4).
\label{ZGammaLRRI'nondiag}
\ee 
Since the conversion factors are diagonal, the one-loop nondiagonal elements of the RI$'$ renormalization factors are equal to the corresponding $\MSbar$ expressions.

\section{Extension to general Wilson-line lattice operators with ${\normalsize n}$ cusps}
\label{sec.IV}

The current study of staple operators, along with our previous work on straight-line operators \cite{Constantinou:2017sej}, lead us to some interesting conclusions about nonlocal operators. From these two cases, we can completely deduce the renormalization coefficients of a general Wilson-line operator with $n$ cusps, defined on the lattice; in particular, we determine both the divergent (linear and logarithmic) and the finite parts of multiplicative renormalizations, as well as all mixing coefficients. We can also justify the nature of the mixing in each case.

\bigskip

All the above coefficients can be deduced from the difference between the bare Green's functions on the lattice and the corresponding $\MSbar$-renormalized Green's functions, obtained in DR: ${\Delta \Lambda}_\Gamma \equiv \Lambda_\Gamma^{\rm LR} - \Lambda_\Gamma^\MSbar$. Below we have gathered our results for these differences, in the case of both straight-line (Ref. \cite{Constantinou:2017sej}) and staple (this work) operators, presented separately for each Feynman diagram, \\

\hspace{-0.5cm}\underline{Straight-line operators,}
\be 
{\left(\Lambda_\Gamma^{\rm straight}\right)}^{\rm LR}_{d_i} - {\left(\Lambda_\Gamma^{\rm straight}\right)}^{\MSbar}_{d_i} \equiv - \frac{g^2 C_F}{16 \pi^2} \ e^{i q_{\mu_1} z} \cdot \mathcal{F}_{d_i}^{\text{straight}} + \mathcal{O} (g^4), \qquad (i=1,2,3,4), \label{Lambda_straight}
\ee
where
\be
\mathcal{F}_{d_1}^{\text{straight}} = 0, \vspace{1.5mm}
\ee 
\be 
\mathcal{F}_{d_2 + d_3}^{\text{straight}} = 2 \ \Gamma \left[ \alpha_5 + 3.7920 \, \beta + (1 - \beta) \log (a^2 \bar{\mu}^2)\right] + \text{sgn}(z) (\Gamma \gamma_{\mu_1} + \gamma_{\mu_1} \Gamma) (\alpha_3 + \alpha_4 c_{SW}), \label{d2d3straight}
\ee
\be
\mathcal{F}_{d_4}^{\text{straight}} = \Gamma \left[ \alpha_6 - 3.7920 \, \beta + (2 + \beta) \log (a^2 \bar{\mu}^2) + \alpha_2 \frac{| z |}{a} \right]. \label{d4straight}
\ee \\
\underline{Staple operators,}
\be 
{\left(\Lambda_\Gamma^{\rm staple}\right)}^{\rm LR}_{d_i} - {\left(\Lambda_\Gamma^{\rm staple}\right)}^{\MSbar}_{d_i} \equiv - \frac{g^2 C_F}{16 \pi^2} \ e^{i q_{\mu_1} z} \cdot \mathcal{F}_{d_i}^{\text{staple}} + \mathcal{O} (g^4), \qquad (i=1,2,3,4),
\ee
where
\be
\mathcal{F}_{d_1}^{\text{staple}} = 0, \vspace{1.5mm}
\ee
\be
\mathcal{F}_{d_2 + d_3}^{\text{staple}} = 2 \ \Gamma \left[ \alpha_5 + 3.7920 \, \beta + (1 - \beta) \log (a^2 \bar{\mu}^2)\right] + \text{sgn}(y) (\Gamma \gamma_{\mu_2} - \gamma_{\mu_2} \Gamma) (\alpha_3 + \alpha_4 c_{SW}), \label{d2d3staple}
\ee
\begin{gather}
\mathcal{F}_{d_4}^{\text{staple}} = \Gamma\Big\{ 3 \left[\alpha_6 - 3.7920 \, \beta + (2 + \beta) \log (a^2 \bar{\mu}^2)\right] + \alpha_2 \frac{| z | + 2 | y |}{a} \nonumber \\
\hspace{6cm}+ 2 \left[\alpha_7 + 3.7920 \, \beta - \beta \log (a^2 \bar{\mu}^2)\right] \Big\}. \label{d4staple}
\end{gather}
The coefficients $\alpha_i$ are numerical constants, which depend on the Symanzik coefficients of the gluon action in use; their values for Wilson, tree-level Symanzik and Iwasaki gluons are given in Tables \ref{tab:stapleLR} and \ref{tab:alpha}. 
\begin{table}[thb!]
  \centering
  \begin{tabular}{|l|l|l|l|l|l|l|}
  \hline
\ \textbf{Gluon action} & \quad \ $\boldsymbol{\alpha_5}$ & \ \quad $\boldsymbol{\alpha_6}$ & \ \quad $\boldsymbol{\alpha_7}$  \\
\hline
\hline
\ Wilson & \ -4.4641 \ & \ -4.5258 \ & \ \ \quad 0 \ \\
\ Tree-level Symanzik \ & \ -4.3413 \ & \ -3.9303 \ & \ -0.8099 \ \\ 
\ Iwasaki & \ -4.1637 \ & \ -1.9053 \ & \ -2.1011 \ \\
\hline
  \end{tabular}
  \caption{Numerical values of the coefficients $\alpha_5{-}\alpha_7$ appearing in the one-loop bare lattice Green's functions of Wilson-line operators (straight line and staple).}
  \label{tab:alpha}
\end{table}
Comparing the above results for the two types of operators, we come to the following conclusions which can be generalized to Wilson-line lattice operators of arbitrary shape:
\begin{itemize}
\item The linear divergence [$\mathcal{O} (1/a)$] depends on the Wilson line's length. 
\item Diagram $d_1$ gives a finite, regulator-independent result in all cases.
\item The only contribution of sail diagrams ($d_2$ and $d_3$) to ${\Delta \Lambda}_\Gamma$ comes from their end points. This is because any parts of a segment which do not include the end points will give finite contributions to $\Lambda_\Gamma^{\rm LR}$, in which the na{\"i}ve continuum limit $a \rightarrow 0$ can be taken, leading to the same result as in DR and thus to a vanishing contribution in ${\Delta \Lambda}_\Gamma$. Consequently the shape of the Wilson line is largely irrelevant and, indeed, all numerical coefficients in Eq. \eqref{d2d3staple} coincide with those in Eq. \eqref{d2d3straight}. The only dependence on the shape regards the Dirac structure of the operator which mixes with $\mathcal{O}_\Gamma$. The mixing terms depend on the direction of the Wilson line in the end points. For the straight Wilson line, the direction in both end points is ${\rm sgn(z)} \hat{\mu}_1$, which leads to the appearance of the additional Dirac structure ${\rm sgn(z)} (\Gamma \gamma_{\mu_1} + \gamma_{\mu_1} \Gamma)$ upon adding together sail diagrams $d_2$ and $d_3$. For the staple Wilson line, the direction in the left end point is ${\rm sgn(y)} \hat{\mu}_2$ and in the right end point is $- {\rm sgn(y)} \hat{\mu}_2$; thus, the additional Dirac structure which appears upon adding the two sail diagrams is $ {\rm sgn(y)} (\Gamma \gamma_{\mu_2} - \gamma_{\mu_2} \Gamma)$. 

\bigskip

The mixing pairs for each type of nonlocal operator can be also explained (partially) by symmetry arguments. For straight-line operators, there is a residual rotational (or hypercubic, on the lattice) symmetry (including reflections) with respect to the three transverse directions to the $\hat{\mu}$ direction parallel to the Wilson line. As a consequence, operators which transform in the same way under this residual symmetry can mix among themselves, under renormalization; i.e., mixing can occur only among the pairs of operators ($\mathcal{O}_\Gamma$, $\mathcal{O}_{\Gamma \gamma_{\mu}}$). This argument can now be applied to a general Wilson line: given that only end points contribute, mixing can occur only with $\mathcal{O}_{\Gamma \gamma_\mu}$, where $\hat{\mu}$ refers to the directions of the two end points of the line. Clearly, the subsets of operators which finally mix depend on the commutation properties between $\Gamma$ and $\gamma_\mu$. We note that, if the fermion action in use preserves chiral symmetry, then none of the operators will mix with each other. 
\item The tadpole diagram ($d_4$) for the staple operators gives, aside from the linearly divergent terms, two types of contributions: one corresponds to each of the three straight-line segments [first square bracket in Eq. \eqref{d4staple}], which is identical to the corresponding contribution from the straight Wilson line, multiplied by a factor of 3, and another contribution for each of the two cusps [second square bracket in Eq. \eqref{d4staple}], multiplied by a factor of 2, which cannot be obtained from the study of straight-line operators.
\end{itemize}
As a consequence of the above, it follows that the difference $\Lambda_\Gamma^{\rm LR} - \Lambda_\Gamma^{\MSbar}$ for a general Wilson-line operator with $n$ cusps (and, hence, $n+1$ segments), defined on the lattice, can be fully extracted from the combination of our results for the straight-line and the staple operators: the contributions of each straight-line segment and each cusp, appearing in the general operators, are obtained from Eqs. (\ref{Lambda_straight} -- \ref{d4staple}). Therefore, without performing any new calculations, the result for the Green's functions of general Wilson-line lattice operators with $n$ cusps, is determined below,
\be 
{\left(\Lambda_\Gamma^{\rm general}\right)}^{\rm LR}_{d_i} - {\left(\Lambda_\Gamma^{\rm general}\right)}^{\MSbar}_{d_i} \equiv - \frac{g^2 C_F}{16 \pi^2} \ e^{i q_{\mu_1} z} \cdot \mathcal{F}_{d_i}^{\text{general}} + \mathcal{O} (g^4), \qquad (i=1,2,3,4),
\ee
where
\be
\mathcal{F}_{d_1}^{\text{general}} = 0, \vspace{1.5mm}
\ee
\be 
\mathcal{F}_{d_2 + d_3}^{\text{general}} = 2 \ \Gamma \left[ \alpha_5 + 3.7920 \, \beta + (1 - \beta) \log (a^2 \bar{\mu}^2)\right] + (\Gamma \hat{\slashed{\mu}}_i + \hat{\slashed{\mu}}_{\hspace{-0.4mm}f} \Gamma) (\alpha_3 + \alpha_4 c_{SW}),
\ee
\be 
\mathcal{F}_{d_4}^{\text{general}} = \Gamma \Big\{ (n + 1) \alpha_6 - 3.7920 \, \beta + \left[ 2 (n + 1) + \beta \right] \log (a^2 \bar{\mu}^2) + \frac{L}{a} \alpha_2 + n \alpha_7 \Big\},
\ee  
$L$ is the Wilson line's length and $\hat{\mu}_{_i} \ (\hat{\mu}_{_{\hspace{-0.5mm}f}})$ is the direction of the Wilson line in the initial (final) end point. In the above relations, it is explicit that there is mixing between the pairs of operators $(\mathcal{O}_\Gamma, \mathcal{O}_{\Gamma \hat{\slashed{\mu}}_i + \hat{\slashed{\mu}}_{\hspace{-0.4mm}f} \Gamma})$. Proceeding further with the renormalization of these operators, we extract the renormalization factors in the $\MSbar$ scheme,
\be 
Z_{\Gamma (diag.)}^{{\rm LR}, \MSbar} = 1 + \frac{g^2 C_F}{16 \pi^2} \Big[ e^\Gamma - \alpha_2 \frac{L}{a} + e_2^\psi c_{SW} + e_3^\psi c_{SW}^2 - (2n + 3) \log \left(a^2 \bar\mu^2\right) \Big] + \mathcal{O} (g^4),
\label{ZGammageneraldiag}
\ee 
\be 
Z_{\Gamma (nondiag./mix.)}^{{\rm LR}, \MSbar} = \frac{g^2 C_F}{16 \pi^2} (-1) \Big[ \alpha_3 + \alpha_4 c_{SW} \Big] + \mathcal{O} (g^4),
\label{ZGammageneralnondiag}
\ee
where $e^\Gamma = \left[e_1^\psi + 1 - 2 \alpha_5 - (n+1) \alpha_6 - n \alpha_7\right]$ and $e_i^\psi$ are given in Appendix \ref{ap.B}. It is worth noting that the results in Eqs. (\ref{ZGammageneraldiag}, \ref{ZGammageneralnondiag}) are both gauge invariant, as was expected.

\section{Conclusions and Future Plans}
\label{sec.V}

In this paper, we have studied the one-loop renormalization of the nonlocal staple-shaped Wilson-line quark operators, both in dimensional regularization (DR) and on the lattice (Wilson/clover massless fermions and Symanzik-improved gluons). This is a follow-up calculation of Ref. \cite{Constantinou:2017sej}, in which straight-line nonlocal operators are studied. These perturbative studies are parts of a wider community effort for investigating the renormalization of nonlocal operators employed in lattice computations of parton distributions (PDFs, GPDs, TMDs) of hadronic physics. A novel aspect of this calculation is the presence of cusps in the Wilson line included in the definition of the nonlocal operators under study, which results in the appearance of additional logarithmic divergences. Perturbative studies of such nonsmooth operators had not been carried out previously on the lattice. As in the case of the straight-line operators, certain pairs of these nonlocal operators mix under renormalization, for chirality-breaking lattice actions, such as the Wilson/clover fermion action. The path structure of each type of nonlocal operator (straight-line, staple, \ldots) leads to different mixing pairs. The results of the present study provide additional information on the renormalization of general nonlocal operators on the lattice.

\bigskip

Particular novel outcomes of our calculation are
\begin{itemize}
\item The one-loop results for the amputated two-point one-particle-irreducible (1-PI) Green’s functions of the staple operators both in DR [Eqs. (\ref{Eq. A1} -- \ref{Eq. A10})] and on the lattice [Eqs. (\ref{Lambda_LR1}, \ref{Lambda_LR2})].
\item The mixing pairs of the staple operators: $(\mathcal{O}_P, \mathcal{O}_{A_{\mu_2}}), (\mathcal{O}_{V_i}, \mathcal{O}_{T_{i\mu_2}}), \ i {\neq} \mu_2$ (for notation, see Sec. \ref{operatorsetup}). We propose a minimal RI$'$-like condition [Eqs. (\ref{RC1} -- \ref{RC3})], which disentangles this mixing and which is appropriate for nonperturbative calculations of parton-distribution functions on the lattice.
\item The one-loop expressions for the renormalization factors of the staple operators in both dimensional and lattice regularizations, in the $\MSbar$ scheme and the proposed RI$'$ scheme [Eqs. (\ref{ZGammaDRMSbar}, \ref{ZGammaDRRI'}, \ref{ZGammaLRMSbar} -- \ref{ZGammaLRRI'nondiag})].
\item The one-loop conversion factors between the RI$'$ and $\MSbar$ schemes [Eqs. (\ref{C_S} -- \ref{C_rest})].
\item An extension of our calculations to general Wilson-line lattice operators with $n$ cusps; we have provided results for their renormalization factors [Eqs. (\ref{ZGammageneraldiag}, \ref{ZGammageneralnondiag})]. 
\end{itemize}

\bigskip

Our results are useful for improving the nonperturbative investigations of transverse momentum-dependent distribution functions (TMDs) on the lattice. Such an example is the calculation of the generalized $g_{1T}$ worm-gear shift in the TMD limit ($|\eta|{\to}\infty$); this quantity involves a ratio between the axial and vector operators. A recent study of TMDs on the lattice \cite{Yoon:2017qzo} reveals tension between results for $g_{1T}$ in the clover and domain-wall formulations. This is not observed in other structures and is an indication of nonmultiplicative renormalization. Our proposed RI$'$-type scheme can be applied to the nonperturbative evaluation of renormalization factors and mixing coefficients of the unpolarized, helicity and transversity quasi-TMDs; this is expected to fix the inconsistency between the two calculations of $g_{1T}$. Also, our one-loop conversion factors can be used to convert the RI$'$ nonperturbative results to the $\MSbar$ scheme. Our results for general Wilson-line lattice operators with $n$ cusps can be used in the nonperturbative renormalization of more general continuum nonlocal operators. 

\bigskip

Comparing our results for the staple operators with the corresponding ones for the straight-line operators, we deduce that the strength of the linear divergences is the same for both types of operators; the presence of cusps lead to additional logarithmic divergences in the staple operators. Also, the observed mixing pairs among operators with different Dirac structures depend on the direction of Wilson line in the end points, and thus, they are different between the two types of operators: the straight-line operator $\mathcal{O}_\Gamma$ mixes with $\mathcal{O}_{\{\Gamma, \gamma_{\mu_1} \}}$, while the staple operator $\mathcal{O}_\Gamma$ mixes with $\mathcal{O}_{[\Gamma, \gamma_{\mu_2} ]}$ (for notation, see Sec. \ref{operatorsetup}). However, the values of the mixing coefficients are the same in the two cases. 

\bigskip

Further perturbative investigations of the staple operators can lead to improved and more robust results. Our future plans include three extensions of the present calculation:
\begin{itemize}
\item The first one is the one-loop evaluation of lattice artifacts to all orders in the lattice spacing, for a range of numerical values of the external quark momentum, of the momentum renormalization scales, and of the action parameters, which are mostly used in simulations. Such a procedure has been successfully employed to local operators \cite{Constantinou:2010gr, Constantinou:2014fka, Alexandrou:2015sea}. The subtraction of the unwanted contributions of the finite lattice spacing from the nonperturbative estimates is essential in order to reduce large cutoff effects in the renormalized Green's functions of the operators and to guarantee a rapid convergence to the continuum limit.
\item Secondly, we intend to add stout smearing on gluon links appearing in the definition of the staple operators and to investigate its impact to the elimination of ultraviolet (UV) divergences and of operator mixing; modern simulations employ such smearing techniques for more convergent results.
\item Thirdly, a natural continuation of the present work is the two-loop calculation of the conversion factors between the RI$'$ and $\MSbar$ schemes; higher-loop corrections will eliminate large truncation effects from the nonperturbative results. Based on our extensive studies for systematic uncertainties on the renormalization functions for the straight Wilson line \cite{Alexandrou:2017huk,Alexandrou:2019lfo}, we find empirically that the one-loop conversion factor is sufficient for lattice spacing satisfying $|z|/a \leq 7{-}8$ and ${(a \mu)}^2$ within the interval $[2-4]$. Outside these regions, a two-loop conversion factor would be called for; clearly, however, other systematic uncertainties will also become more relevant (lattice artifacts, volume effects, etc).  
\end{itemize}
Finally, our perturbative analysis can be also applied to the study of further composite Wilson-line operators, relevant to different quasidistribution functions, e.g., gluon quasi-PDFs, etc.

\vspace*{1cm}
\centerline{{\bf\large{{\bf{Acknowledgements}}}}}
\bigskip
M.C. acknowledges financial support by the U.S. Department of Energy, Office of Nuclear Physics, within the framework of the TMD Topical Collaboration, as well as, by the National Science Foundation under Grant No. PHY-1714407. G.S. acknowledges financial support by the University of Cyprus, within the framework of Ph.D. student scholarships.

\appendix
\section{Green's functions in dimensional regularization}
\label{ap.A}

In this appendix, the full expressions for the one-loop amputated Green's functions of the staple operators $\Lambda_\Gamma^{\rm 1-loop}$, calculated in dimensional regularization (DR), are presented in a compact form [Eqs. (\ref{Eq. A1} -- \ref{Eq. A10})]. From these expressions it is straightforward to derive the renormalized Green's functions, both in the $\MSbar$ scheme [by removing the $\mathcal{O}(1/\varepsilon)$ terms] and in any variant of the RI$'$ scheme, as described in Sec. \ref{sec.IIC}; the corresponding conversion factors [Eqs. (\ref{CGamma1} -- \ref{CGamma3})] also follow immediately. The functions $\Lambda_\Gamma^{\rm 1-loop}$ depend on integrals of modified Bessel functions of the second kind, $K_n$, over Feynman parameters and/or over $\zeta$-variables stemming from the definition of the staple operators. These integrals are denoted by $P_i \equiv P_i (q^2, q_{\mu_1}, z)$, $Q_i \equiv Q_i (q^2, q_{\mu_1}, q_{\mu_2}, z, y)$ and $R_i \equiv R_i (q^2, q_{\mu_1}, q_{\mu_2}, z, y)$; they are listed at the end of this appendix [Eqs. (\ref{P1} -- \ref{R8})].

\begin{align}
\Lambda_S^{\text{1-loop}} = \frac{g^2 C_F}{16 \pi^2} \boldmath\Bigg\{& \Lambda_S^{\text{tree}} \pmb{\bigg[} 
\begin{aligned}[t]
&(8-\beta) \left(2 + \frac{1}{\varepsilon} + \log \left(\frac{\overline{\mu}^2}{q^2}\right) \right) +2 (\beta + 6) \gamma_E +(\beta +2) \log\left(\frac{q^2 z^2}{4}\right) \nonumber \\
&+4 \log \left(\frac{q^2 y^2}{4}\right) +4 \left(2 \ \frac{y}{z} \tan^{-1}\left(\frac{y}{z}\right)-\log\left(1+\frac{y^2}{z^2}\right)\right) + 2 (\beta +2) P_1 \nonumber \\
&-2 \beta \sqrt{q^2} \left| z\right| P_4 +2 q_{\mu_1} \bigg(2 Q_4-i \Big(2 Q_1-\beta  z (P_1-P_2)\Big)\bigg) \nonumber \\
&+4 q_{\mu_2} \Big(R_6 - R_2 + i (R_1-R_4)\Big) \pmb{\bigg]} \nonumber 
\end{aligned} \\
&+ \Lambda_{T_{\mu_1 \mu_2}}^{\text{tree}} \pmb{\bigg[}4 \sqrt{q^2}(y Q_3 + z R_5)\pmb{\bigg]} + \Lambda_{V_{\mu_1}}^{\text{tree}} \slashed{q} \ \pmb{\bigg[}-4 Q_4 \pmb{\bigg]} +\Lambda_{V_{\mu_2}}^{\text{tree}} \slashed{q} \ \pmb{\bigg[}4 i (R_4-R_1)\pmb{\bigg]}\boldmath\Bigg\}, 
\label{Eq. A1}
\end{align}
\begin{align}
\Lambda_P^{\text{1-loop}} = \gamma_5 \Lambda_S^{\text{1-loop}},
\end{align}
\begin{align}
\Lambda_{V_{\mu_1}}^{\text{1-loop}} = \frac{g^2 C_F}{16 \pi^2} \boldmath\Bigg\{& \Lambda_{V_{\mu_1}}^{\text{tree}} \pmb{\bigg[} 
\begin{aligned}[t]
&\beta + (8-\beta) \left(2 + \frac{1}{\varepsilon} + \log \left(\frac{\overline{\mu}^2}{q^2}\right) \right) +2 (\beta + 6) \gamma_E +(\beta +2) \log\left(\frac{q^2 z^2}{4}\right) \nonumber \\
&+4 \log \left(\frac{q^2 y^2}{4}\right) +4 \left(2 \ \frac{y}{z} \tan^{-1}\left(\frac{y}{z}\right)-\log\left(1+\frac{y^2}{z^2}\right)\right) +2 (\beta  P_1-2 P_2) \nonumber \\
&-\frac{1}{2} \beta  q^2 z^2 (P_1-P_2) -2 i q_{\mu_1} \Big(2 (Q_1+Q_2+z P_3) - \beta  z (P_1-P_2)\Big) \nonumber \\
&-2 \sqrt{q^2} \left| z\right|  (\beta  P_4-2 P_5) + 4 q_{\mu_2} \Big(R_6 - R_2 + i (R_1-R_4)\Big)\pmb{\bigg]} \nonumber 
\end{aligned} \\
&+ \Lambda_{V_{\mu_2}}^{\text{tree}} \pmb{\bigg[} 4 i \bigg( \sqrt{q^2} (y Q_5 + z R_8) + q_{\mu_1} \Big(R_4-R_1 + i (R_3-R_7)\Big) \bigg) \pmb{\bigg]} \nonumber \\
&+ \Lambda_S^{\text{tree}} \slashed{q} \pmb{\bigg[}
\begin{aligned}[t]
& i \Big(4 (Q_2-z P_3)-\beta  \sqrt{q^2} z \left| z\right|  (P_4-P_5)\Big) \nonumber \\
&+ q_{\mu_1} \bigg(2 \frac{\left| z\right|}{\sqrt{q^2}} \Big((\beta -2) P_4+2 P_5\Big) + \beta z^2 P_3 \bigg) \pmb{\bigg]} \nonumber 
\end{aligned} \\
&+ \Lambda_{T_{\mu_1 \mu_2}}^{\text{tree}} \slashed{q} \pmb{\bigg[} 4 i (R_4-R_1)\pmb{\bigg]}\boldmath\Bigg\},
\end{align}
\begin{align}
\Lambda_{V_{\mu_2}}^{\text{1-loop}} = \frac{g^2 C_F}{16 \pi^2} \boldmath\Bigg\{& \Lambda_{V_{\mu_2}}^{\text{tree}} \pmb{\bigg[} 
\begin{aligned}[t]
&(8-\beta) \left(2 + \frac{1}{\varepsilon} + \log \left(\frac{\overline{\mu}^2}{q^2}\right) \right) +2 (\beta + 6) \gamma_E +(\beta +2) \log\left(\frac{q^2 z^2}{4}\right) \nonumber \\
&+4 \log \left(\frac{q^2 y^2}{4}\right) +4 \left(2 \ \frac{y}{z} \tan^{-1}\left(\frac{y}{z}\right)-\log\left(1+\frac{y^2}{z^2}\right)\right) +2 (\beta  P_1-2 P_2) \nonumber \\
&-\beta \sqrt{q^2} \left| z\right| P_4 +2 q_{\mu_1} \bigg(2 Q_4-i \Big(2 Q_1-\beta  z (P_1-P_2)\Big)\bigg) \nonumber \\
&-4 q_{\mu_2} (R_2+R_3-R_6-R_7) \pmb{\bigg]} \nonumber 
\end{aligned} \\
&+\Lambda_{V_{\mu_1}}^{\text{tree}} \pmb{\bigg[} 
\begin{aligned}[t]
&-i \bigg( 4 \sqrt{q^2} (y Q_5 + z R_8) \nonumber \\
&+ q_{\mu_2} \left(4 (Q_2 + z P_3 - i Q_4) -\beta  \sqrt{q^2} z \left| z\right| (P_4-P_5)\right) \bigg) \pmb{\bigg]} \nonumber
\end{aligned} \\
&+ \Lambda_S^{\text{tree}} \slashed{q} \pmb{\bigg[} 4 (R_3-R_7) + q_{\mu_2} \bigg(2 \frac{\left| z\right|}{\sqrt{q^2}} \Big((\beta -2) P_4+2 P_5\Big) - \beta z^2 P_3 \bigg) \pmb{\bigg]} \nonumber \\
&+ \Lambda_{T_{\mu_1 \mu_2}}^{\text{tree}} \slashed{q} \pmb{\bigg[} 4 Q_4 \pmb{\bigg]}\boldmath\Bigg\},
\end{align}
\begin{align}
\Lambda_{V_{\nu}}^{\text{1-loop}} = \frac{g^2 C_F}{16 \pi^2} \boldmath\Bigg\{& \Lambda_{V_{\nu}}^{\text{tree}} \pmb{\bigg[} 
\begin{aligned}[t]
&(8-\beta) \left(2 + \frac{1}{\varepsilon} + \log \left(\frac{\overline{\mu}^2}{q^2}\right) \right) +2 (\beta + 6) \gamma_E +(\beta +2) \log\left(\frac{q^2 z^2}{4}\right) \nonumber \\
&+4 \log \left(\frac{q^2 y^2}{4}\right) +4 \left(2 \ \frac{y}{z} \tan^{-1}\left(\frac{y}{z}\right)-\log\left(1+\frac{y^2}{z^2}\right)\right) + 2 (\beta  P_1-2 P_2) \nonumber \\
&- \beta \sqrt{q^2} \left| z\right| P_4 +2 q_{\mu_1} \bigg(2 Q_4-i \Big(2 Q_1-\beta  z (P_1-P_2)\Big)\bigg) \nonumber \\
&+4 q_{\mu_2} \Big(R_6 - R_2 + i (R_1-R_4)\Big) \pmb{\bigg]} \nonumber 
\end{aligned} \\
&+\Lambda_{V_{\mu_1}}^{\text{tree}} q_\nu \pmb{\bigg[} -i \left(4 (Q_2 + z P_3 - i Q_4) -\beta  \sqrt{q^2} \ z \left| z\right| (P_4-P_5)\right) \pmb{\bigg]} \nonumber \\
&+ \Lambda_{V_{\mu_2}}^{\text{tree}} q_{\nu} \pmb{\bigg[} 4 i \Big(R_4-R_1 + i (R_3-R_7)\Big) \pmb{\bigg]} + \varepsilon_{\mu_1 \mu_2 \nu \rho} \ \Lambda_{A_{\rho}}^{\text{tree}} \pmb{\bigg[} 4 \sqrt{q^2} (y Q_3 + z R_5) \pmb{\bigg]} \nonumber \\
&+ \Lambda_S^{\text{tree}} \slashed{q} q_{\nu} \pmb{\bigg[} 2 \frac{\left| z\right|}{\sqrt{q^2}} \Big((\beta -2) P_4+2 P_5\Big) - \beta z^2 P_3 \pmb{\bigg]} + \Lambda_{T_{\mu_1 \nu}}^{\text{tree}} \slashed{q} \pmb{\bigg[} 4 Q_4 \pmb{\bigg]} \nonumber \\
&+ \Lambda_{T_{\mu_2 \nu}}^{\text{tree}} \slashed{q} \pmb{\bigg[} 4 i (R_1-R_4)\pmb{\bigg]}\boldmath\Bigg\}, \qquad (\nu \neq \mu_1, \mu_2)
\end{align}
\begin{align}  
\Lambda_{A_{\mu_1}}^{\text{1-loop}} = \gamma_5 \Lambda_{V_{\mu_1}}^{\text{1-loop}}, \quad \Lambda_{A_{\mu_2}}^{\text{1-loop}} = \gamma_5 \Lambda_{V_{\mu_2}}^{\text{1-loop}} \quad \Lambda_{A_{\nu}}^{\text{1-loop}} = \gamma_5 \Lambda_{V_{\nu}}^{\text{1-loop}}, \qquad (\nu \neq \mu_1, \mu_2)
\end{align}
\begin{align}
\Lambda_{T_{\mu_1 \mu_2}}^{\text{1-loop}} = \frac{g^2 C_F}{16 \pi^2} \boldmath\Bigg\{& \Lambda_{T_{\mu_1 \mu_2}}^{\text{tree}} \pmb{\bigg[} 
\begin{aligned}[t]
&\beta + (8-\beta) \left(2 + \frac{1}{\varepsilon} + \log \left(\frac{\overline{\mu}^2}{q^2}\right) \right) +2 (\beta + 6) \gamma_E +(\beta +2) \log\left(\frac{q^2 z^2}{4}\right) \nonumber \\
&+4 \log \left(\frac{q^2 y^2}{4}\right) +4 \left(2 \ \frac{y}{z} \tan^{-1}\left(\frac{y}{z}\right)-\log\left(1+\frac{y^2}{z^2}\right)\right) +2 (\beta - 2) P_1 \nonumber \\
&-\frac{1}{2} \beta  q^2 z^2 (P_1-P_2) -2 i q_{\mu_1} \Big(2 (Q_1+Q_2) - \beta  z (P_1-P_2)\Big) \nonumber \\
&-\beta \sqrt{q^2} \left| z\right| P_4 -4 q_{\mu_2} (R_2+R_3-R_6-R_7) \pmb{\bigg]} \nonumber 
\end{aligned} \\
&+ \Lambda_S^{\text{tree}} \pmb{\bigg[} 
\begin{aligned}[t]
&-4 \sqrt{q^2}(y Q_3 + z R_5) +4 i q_{\mu_1} \Big(R_4-R_1 + i (R_3-R_7)\Big)  \nonumber \\
&+ i q_{\mu_2} \left(4 (Q_2 - i Q_4) -\beta  \sqrt{q^2} \ z \left| z\right| (P_4-P_5)\right) \pmb{\bigg]}\nonumber
\end{aligned} \\
&+ \Lambda_{V_{\mu_1}}^{\text{tree}} \slashed{q} \pmb{\bigg[} 4 (R_3-R_7) + \beta q_{\mu_2} z^2 P_3 \pmb{\bigg]} \nonumber \\
&+ \Lambda_{V_{\mu_2}}^{\text{tree}} \slashed{q} \pmb{\bigg[} -i \left(4 Q_2 - \beta \sqrt{q^2} \ z \left| z\right| (P_4-P_5)\right)-\beta q_{\mu_1} z^2 P_3 \pmb{\bigg]} \boldmath\Bigg\},
\end{align}
\begin{align}
\Lambda_{T_{\mu_1 \nu}}^{\text{1-loop}} = \frac{g^2 C_F}{16 \pi^2} \boldmath\Bigg\{& \Lambda_{T_{\mu_1 \nu}}^{\text{tree}} \pmb{\bigg[} 
\begin{aligned}[t]
&\beta + (8-\beta) \left(2 + \frac{1}{\varepsilon} + \log \left(\frac{\overline{\mu}^2}{q^2}\right) \right) +2 (\beta + 6) \gamma_E +(\beta +2) \log\left(\frac{q^2 z^2}{4}\right) \nonumber \\
&+4 \log \left(\frac{q^2 y^2}{4}\right) +4 \left(2 \ \frac{y}{z} \tan^{-1}\left(\frac{y}{z}\right)-\log\left(1+\frac{y^2}{z^2}\right)\right) +2 (\beta - 2) P_1 \nonumber \\
&-\frac{1}{2} \beta  q^2 z^2 (P_1-P_2) -2 i q_{\mu_1} \Big(2 (Q_1+Q_2) - \beta  z (P_1-P_2)\Big) \nonumber \\
&-\beta \sqrt{q^2} \left| z\right| P_4 +4 q_{\mu_2} \Big(R_6 - R_2 + i (R_1-R_4)\Big) \pmb{\bigg]} \nonumber 
\end{aligned} \\
&+ \Lambda_{T_{\mu_1 \mu_2}}^{\text{tree}} \pmb{\bigg[} 4 i q_{\nu} (R_4-R_1+i (R_3-R_7)) \pmb{\bigg]} \nonumber \\
&+ \Lambda_{T_{\mu_2 \nu}}^{\text{tree}} \pmb{\bigg[} 4 i \left(\sqrt{q^2} (y Q_5 +z R_8) + q_{\mu_1} (R_4-R_1+i (R_3-R_7))\right) \pmb{\bigg]} \nonumber \\
&+ \Lambda_S^{\text{tree}} q_{\nu} \pmb{\bigg[} i \left(4 (Q_2 - i Q_4) -\beta  \sqrt{q^2} \ z \left| z\right| (P_4-P_5)\right) \pmb{\bigg]} +\Lambda_{V_{\mu_1}}^{\text{tree}} \slashed{q} \pmb{\bigg[} \beta q_{\nu} z^2 P_3 \pmb{\bigg]}\nonumber \\
&+ \Lambda_{V_{\nu}}^{\text{tree}} \slashed{q} \pmb{\bigg[} -i \left(4 Q_2 - \beta \sqrt{q^2} \ z \left| z\right| (P_4-P_5)\right)-\beta q_{\mu_1} z^2 P_3 \pmb{\bigg]} \nonumber \\
&+ \epsilon_{\mu_1 \mu_2 \nu \rho} \ \Lambda_{A_{\rho}}^{\text{tree}} \slashed{q} \pmb{\bigg[} 4 i (R_1-R_4) \pmb{\bigg]} \boldmath\Bigg\}, \qquad (\nu \neq \mu_1, \mu_2)
\end{align}
\begin{align}
\Lambda_{T_{\mu_2 \nu}}^{\text{1-loop}} = - \gamma_5 \ \varepsilon_{\mu_1 \mu_2 \nu \rho} \ \Lambda_{T_{\mu_1 \rho}}^{\text{1-loop}}, \qquad (\nu \neq \mu_1, \mu_2)
\label{Eq. A9}
\end{align}
\begin{align}
\Lambda_{T_{\nu \rho}}^{\text{1-loop}} = - \gamma_5 \ \varepsilon_{\mu_1 \mu_2 \nu \rho} \ \Lambda_{T_{\mu_1 \mu_2}}^{\text{1-loop}}, \qquad (\nu, \rho \neq \mu_1, \mu_2).
\label{Eq. A10}
\end{align}
In Eqs. (\ref{Eq. A9}, \ref{Eq. A10}), $\varepsilon_{\mu_1 \mu_2 \nu \rho}$ is the Levi-Civita tensor, $\varepsilon_{1234}$ = 1. \\

\textbf{List of integrals:} \\
\\
In what follows, we use the notation: $s \equiv \sqrt{q^2 (1-x) x}$.
\begin{align}
&P_1 (q^2, q_{\mu_1}, z) \equiv \int_0^1 dx \ e^{-i q_{\mu_1} x z} \ K_0\left(s \left| z\right| \right), \label{P1}\\
&P_2 (q^2, q_{\mu_1}, z) \equiv \int_0^1 dx \ e^{-i q_{\mu_1} x z} \ K_0\left(s \left| z\right| \right) x, \\
&P_3 (q^2, q_{\mu_1}, z) \equiv \int_0^1 dx \ e^{-i q_{\mu_1} x z} \ K_0\left(s \left| z\right| \right) x (1-x), \\
&P_4 (q^2, q_{\mu_1}, z) \equiv \int_0^1 dx \ e^{-i q_{\mu_1} x z} \ K_1\left(s \left| z\right| \right) \sqrt{(1-x) x}, \\
&P_5 (q^2, q_{\mu_1}, z) \equiv \int_0^1 dx \ e^{-i q_{\mu_1} x z} \ K_1\left(s \left| z\right| \right) \sqrt{(1-x) x} \ x, 
\end{align}
\begin{align}
&Q_1 (q^2, q_{\mu_1}, q_{\mu_2}, z, y) \equiv \int_0^1 dx \int_0^z d\zeta \ e^{-i q_{\mu_1} x \zeta} \cos (q_{\mu_2} x y) K_0\left(s \sqrt{y^2+\zeta^2}\right), \\
&Q_2 (q^2, q_{\mu_1}, q_{\mu_2}, z, y) \equiv \int_0^1 dx \int_0^z d\zeta \ e^{-i q_{\mu_1} x \zeta} \cos (q_{\mu_2} x y) K_0\left(s \sqrt{y^2+\zeta^2}\right) (1-x), \\
&Q_3 (q^2, q_{\mu_1}, q_{\mu_2}, z, y) \equiv \int_0^1 dx \int_0^z d\zeta \ e^{-i q_{\mu_1} x \zeta} \cos (q_{\mu_2} x y) K_1\left(s \sqrt{y^2+\zeta^2}\right) \frac{\sqrt{(1-x) x}}{\sqrt{y^2+\zeta ^2}}, \\
&Q_4 (q^2, q_{\mu_1}, q_{\mu_2}, z, y) \equiv \int_0^1 dx \int_0^z d\zeta \ e^{-i q_{\mu_1} x \zeta} \sin (q_{\mu_2} x y) K_0\left(s \sqrt{y^2+\zeta^2}\right) (1-x), \\
&Q_5 (q^2, q_{\mu_1}, q_{\mu_2}, z, y) \equiv \int_0^1 dx \int_0^z d\zeta \ e^{-i q_{\mu_1} x \zeta} \sin (q_{\mu_2} x y) K_1\left(s \sqrt{y^2+\zeta^2}\right) \frac{\sqrt{(1-x) x}}{\sqrt{y^2+\zeta ^2}},
\end{align}
\begin{align}
&R_1 (q^2, q_{\mu_2}, y) \equiv \int_0^1 dx \int_0^y d\zeta \ \cos (q_{\mu_2} x \zeta) \  K_0\left(s \left| \zeta\right| \right) (1-x), \\
&R_2 (q^2, q_{\mu_2}, y) \equiv \int_0^1 dx \int_0^y d\zeta \ \sin (q_{\mu_2} x \zeta) \  K_0\left(s \left| \zeta\right| \right), \\
&R_3 (q^2, q_{\mu_2}, y) \equiv \int_0^1 dx \int_0^y d\zeta \ \sin (q_{\mu_2} x \zeta) \  K_0\left(s \left| \zeta\right| \right) (1-x), \\
&R_4 (q^2, q_{\mu_1}, q_{\mu_2}, z, y) \equiv \int_0^1 dx \int_0^y d\zeta \ e^{-i q_{\mu_1} x z} \ \cos (q_{\mu_2} x \zeta) \ K_0\left(s \sqrt{z^2+\zeta^2}\right) (1-x), \\
&R_5 (q^2, q_{\mu_1}, q_{\mu_2}, z, y) \equiv \int_0^1 dx \int_0^y d\zeta \ e^{-i q_{\mu_1} x z} \ \cos (q_{\mu_2} x \zeta) \ K_1\left(s \sqrt{z^2+\zeta^2}\right) \frac{\sqrt{(1-x) x}}{\sqrt{z^2 + \zeta^2}}, \\
&R_6 (q^2, q_{\mu_1}, q_{\mu_2}, z, y) \equiv \int_0^1 dx \int_0^y d\zeta \ e^{-i q_{\mu_1} x z} \ \sin (q_{\mu_2} x \zeta) \ K_0\left(s \sqrt{z^2+\zeta^2}\right), 
\end{align}
\begin{align}
&R_7 (q^2, q_{\mu_1}, q_{\mu_2}, z, y) \equiv \int_0^1 dx \int_0^y d\zeta \ e^{-i q_{\mu_1} x z} \ \sin (q_{\mu_2} x \zeta) \ K_0\left(s \sqrt{z^2+\zeta^2}\right) (1-x), \\
&R_8 (q^2, q_{\mu_1}, q_{\mu_2}, z, y) \equiv \int_0^1 dx \int_0^y d\zeta \ e^{-i q_{\mu_1} x z} \ \sin (q_{\mu_2} x \zeta) \ K_1\left(s \sqrt{z^2+\zeta^2}\right) \frac{\sqrt{(1-x) x}}{\sqrt{z^2 + \zeta^2}}. \label{R8}
\end{align}

\section{Renormalization of fermion fields}
\label{ap.B}

In this appendix, we have gathered useful expressions regarding the renormalization of fermion fields in both dimensional (DR) and lattice (LR) regularizations, taken from, e.g., Refs. \cite{Gracey:2003yr} and \cite{Alexandrou:2012mt}, respectively. We give the one-loop expressions for the renormalization factors in the $\MSbar$ and RI$'$ schemes, as well as the conversion factors between the two schemes,
\begin{align} 
&Z_\psi^{\text{DR},\MSbar} = 1 + \frac{g^2 C_F}{16 \pi^2} (\beta - 1) \frac{1}{\varepsilon} + \mathcal{O} (g^4), \label{ZpsiDRMSbar}\\
&Z_\psi^{\text{DR},\text{RI}'} = 1 + \frac{g^2 C_F}{16 \pi^2} (\beta - 1) \left( \frac{1}{\varepsilon} + 1 +\log \left(\frac{\bar{\mu}^2}{\bar{q}^2}\right) \right) + \mathcal{O} (g^4), \\
&C_\psi^{\text{RI}',\MSbar} = \frac{Z_\psi^{\text{DR},\text{RI}'}}{Z_\psi^{\text{DR},\MSbar}} = \frac{Z_\psi^{\text{LR},\text{RI}'}}{Z_\psi^{\text{LR},\MSbar}} = 1 + \frac{g^2 C_F}{16 \pi^2} (\beta - 1) \left(1 +\log \left(\frac{\bar{\mu}^2}{\bar{q}^2}\right) \right) + \mathcal{O} (g^4), \\
&Z_\psi^{\text{LR},\text{RI}'} = 1 + \frac{g^2 C_F}{16 \pi^2} \left[ e_1^\psi + 4.7920 \, \beta + e_2^\psi c_{SW} + e_3^\psi c_{SW}^2 + (1-\beta) \log \left( a^2 \bar{q}^2 \right) \right] + \mathcal{O} (g^4), \\
&Z_\psi^{\text{LR},\MSbar} = \frac{Z_\psi^{\text{LR},\text{RI}'}}{C_\psi^{\text{RI}',\MSbar}} = 1 + \frac{g^2 C_F}{16 \pi^2} \Big[ (e_1^\psi + 1) + 3.7920 \, \beta + e_2^\psi c_{SW} + e_3^\psi c_{SW}^2 \nonumber \\
& \hspace{9.4cm} + (1-\beta) \log \left( a^2 \bar{\mu}^2 \right) \Big] + \mathcal{O} (g^4). \label{ZpsiLRMSbar}
\end{align}
The numerical constants $e_i^\psi$ depend on the gluon action in use; their values for Wilson, tree-level Symanzik and Iwasaki improved gluon actions are given in Table \ref{tab:Zpsi}.
\begin{table}[thb]
  \centering
  \begin{tabular}{|l|l|l|l|}
  \hline
\ \textbf{Gluon action} & \ \quad \ $\boldsymbol{e_1^\psi}$ & \ \quad $\boldsymbol{e_2^\psi}$ & \ \quad $\boldsymbol{e_3^\psi}$  \\
\hline
\hline
\ Wilson & \ 11.8524 \ & \ -2.2489 \ & \ -1.3973 \ \\
\ Tree-level Symanzik \ & \ \phantom{a}8.2313 \ & \ -2.0154 \ & \ -1.2422 \ \\ 
\ Iwasaki & \ \phantom{a}3.3246 \ & \ -1.6010 \ & \ -0.9732 \ \\
\hline
  \end{tabular}
  \caption{Numerical values of the coefficients $e_1^\psi${-}$e_3^\psi$ appearing in the one-loop renormalization factors of fermion fields on the lattice.}
  \label{tab:Zpsi}
\end{table}

%%%%%%%%%%%%%%%%%%%%%%%%%%%%%%%%%%%%%%%%%%%%%%%%%%%%%%%%%%%%%%%%%%%%
% ========================= REFERENCES =========================
%\newpage
\vspace*{-0.25cm}    
\bibliographystyle{elsarticle-num}                     % Style for bibliography
\bibliography{StapleOperatorsReferences}

%%%%%%%%%%%%%%%%%%%%%%%%%%%%%%%%%%%%%%%%%%%%%%%%%%%%%%%%%%%%%%%%%%%%%%%%%%%%%
\end{document}